\documentclass[preprint2]{aastex}
\usepackage{color}
\usepackage{natbib}
\usepackage{multirow}

\shorttitle{Solar Magnetic Flux Rope Formation during Confined Flaring}
\shortauthors{}

\begin{document}

\title{Formation of Magnetic Flux Ropes during Confined Flaring Well Before the Onset of a Pair of Major Coronal Mass Ejections}

\author{Georgios Chintzoglou}
\affil{School of Physics, Astronomy and Computational Sciences, George Mason University, 
       4400 University Dr., MSN 2F2, Fairfax, VA 22030, USA}
\email{gchintzo@gmu.edu}

\and

\author{Spiros Patsourakos}
\affil{Department of Physics, Section of Astrogeophysics, University of Ioannina, Panepistimioupoli Douroutis, Ioannina, Greece}

\and

\author{Angelos Vourlidas}
\affil{The Johns Hopkins University Applied Physics Laboratory, Laurel, MD 20723, USA}

\begin{abstract}
NOAA Active Region (AR) 11429 was the source of twin super-fast Coronal Mass Ejections (CMEs). The CMEs took place within a hour from each other, with the onset
of the first taking place in the beginning of March 7, 2012.
This AR fulfills all the requirements for a ``super active region''; namely, Hale's law incompatibility and a $\delta$-spot magnetic configuration. 
One of the biggest storms of Solar Cycle 24 to date ($D_{st}=-143$\,nT) was associated with one of these events. Magnetic Flux Ropes (MFRs) are twisted magnetic structures in the corona, best seen in $\sim$10\,MK hot plasma emission and are often considered the core of erupting structures. However, their ``dormant'' existence in the solar atmosphere (i.e. prior to eruptions), is an open question. Aided by multi-wavelength observations (SDO/HMI/AIA and STEREO EUVI B) and a Non-Linear Force-Free (NLFFF) model for the coronal magnetic field, our work uncovers two separate, weakly-twisted magnetic flux systems which suggest the existence of pre-eruption MFRs that eventually became the seeds of the two CMEs. The MFRs could have been formed during confined (i.e. not leading to major CMEs) flaring and sub-flaring events which took place the day before the two CMEs in the host AR 11429.

\end{abstract}

\keywords{Sun: General --- Sun: Surface Magnetism --- Sun: Coronal Mass Ejections --- Sun: Activity --- Sun: Corona --- Sun: Photosphere}

\section{Introduction}\label{INTRO}
Coronal Mass Ejections (CMEs) are the most energetic phenomena of the solar corona. They are explosive ejections of coronal plasma that advect coronal magnetic flux into the heliosphere, which then drives space weather at Earth. We know several things about the structure of the advected magnetic field. About 40\% of CMEs exhibit the structure of a Magnetic Flux-Rope (MFR or flux-rope thereafter) as determined from coronagraphic observations close to the Sun (\citealt{Vourlidas_et_al_2013}). In-situ 1~AU observations suggest that at least a third of CMEs maintain an overall MFR morphology as they propagate into interplanetary space (e.g. \citealt{Kilpua_etal_2013}). Both percentages are lower limits. The MFR morphology can be easily obscured by projection effects in imaging observations. In-situ MFR determinations, or Magnetic Clouds (MCs) in in-situ parlance, are even trickier due to the local nature of in-situ measurement.  Several studies argue that the MCs criteria are too strict and/or the in-situ detectors intercept CMEs far from their MFR core, say close to  the CME legs (\citealt{Kilpua_etal_2013, Demoulin_etal_2013}). In addition, \citet{Riley_Richardson_2013} have  shown that the propensity of a CME to show MFR characteristics decreases with increasing heliocentric distance.
MFRs are magnetic structures comprised of a family of magnetic field lines wrapping around a central field line or axis (hence the term flux-``rope''). The fact that MFRs exhibit a twisted magnetic field line topology is indicative of available free energy and magnetic helicity stored within these stressed magnetic structures.

There is a continuing controversy on the subject of CME initiation. The theories suggest that the core structure of the CME, the MFR, is formed either on-the-fly (e.g., \citealt{Lynch_etal_2008}) or exists \emph{before} the eruption (e.g., \citealt{Kliem_Torok_2006}). The debate translates into whether the physical process behind the eruption is an ideal MHD process (pre-existing MFR), or a non-ideal/resistive process (on-the-fly MFR formation). Further complications arise from the variety of different formation mechanisms of MFRs. These include flux emergence, flux cancellation, magnetic reconnection and shearing (e.g., \citealt{Antiochos_etal_1999}; \citealt{vanBallegooijen_Martens_1989}; \citealt{Yurchyshyn_etal_2006b}; \citealt{Archontis_Torok_2008}; \citealt{Lynch_etal_2008}; \citealt{Archontis_etal_2009}; \citealt{Aulanier_etal_2010}; \citealt{Georgoulis_etal_2012}; \citealt{Leake_etal_2013}; \citealt{Tziotziou_etal_2013}). Moreover, the pre-eruptive structure may be formed at different layers of the solar atmosphere, ranging from the photosphere/lower chromosphere to the corona, or even bodily emerge from the convection zone.

Since the beginning of the SDO era, observational analyses of pre-eruptive configurations are typically case-studies of MFR ``candidates'' seen close to, or at, the solar limb. 
\citet{Cheng_X_etal_P2_2011} have observed indications for the formation of a MFR during the impulsive phase of a CME with a multi-thermal nature, possibly due to new poloidal field added around the MFR via reconnection with the ambient field. \citet{Zhang_etal_2012} provided the first pre-existing MFR observation close to the time (a few minutes) of a CME onset. In this observation a coiled structure, or ``hot channel'' (detected in 94\AA\ \& 131\AA\ EUV emission with plasma temperatures of $\approx$ 6.4 \& 10 MK) was seen to progressively unwind into a semi-circular shape and then erupt as the core of a CME. This observation suggested that an ideal MHD instability, like the Torus Instability, is responsible for the CME eruption. \citet{Cheng_etal_2013} determined that the ``hot channel'' was moving faster than the leading front of a CME, at least in the beginning of the eruption, providing evidence for the role of MFRs as a driver of the CME. \citet{Patsourakos_etal_2013} have provided the first direct evidence of a truly pre-existing MFR. In their observation, the MFR was seen after a confined flare, at $\approx$ 7 hours before the CME. They proposed a scenario based on MFR formation during confined flaring events. A recent statistical study using AIA data showed that hot MFRs are a common occurrence during M and X-class flares, confined or eruptive (\citealt{Nindos_etal_2015}).

Soft X-ray (SXR) observations can also supply important information for MFRs within ARs. \citet{Green_etal_2011} found evidence of flux rope formation by means of photospheric cancellation in an active region in association with the formation of sigmoids seen in the Soft X-Rays (SXR) and in the EUV light. The sigmoidal structures are features which are often seen in SXR prior to the eruption (\citealt{Canfield_etal_1999}; \citealt{McKenzie_Canfield_2008}; \citealt{Savcheva_vanBallego_2009}). \citet{Rust_Kumar_1996} studied a large sample of bright sigmoid structures seen in SXR observations by Yohkoh observatory and they were the first to suggest the possibility that SXR sigmoids might signify the presence of MFRs. \citet{Green_Kliem_2009}, studying the SXR sigmoid topology of an AR, found traces of a bald patch separatrix surface which they interpreted as evidence for the existence of MFR prior to the onset of a CME.

The increasing number of reported MFR ``candidates'' refer to coronal observations away from disk center. The lower background EUV emission, i.e. there are fewer bright loop-like structures along the LOS when MFRs are observed closer to the limb, facilitates the detection of the optically-thin MFRs. But to uncover the magnetic structure,  good knowledge of  the photospheric magnetic field configuration, is necessary. This information allows to, at least, pinpoint the ``roots'' of these line-tied 3D MFR structures. In large viewing angles, photospheric magnetic field measurements suffer from  projection effects (and increased noise) and  yield no reliable information for ARs very close to the limb. Hence, the past studies of the dynamics and formation of MFRs were quite restricted  because they could not make use of coronal magnetic field models, such as the Non-Linear Force-Free Field (NLFF, e.g. \citealt{Canou_etal_2009}; \citealt{Wiegelmann_Sakurai_2012}) or the flux-rope insertion method (e.g. \citealt{vanBallegooijen_2004}; \citealt{Savcheva_vanBallego_2009}; \citealt{Su_etal_2009}; \citealt{Savcheva_etal_2012}; \citealt{Bobra_etal_2008}).  For instance, in the work of \citet{Patsourakos_etal_2013}, the event was at the limb and therefore it was impossible to support the ``confined flare-to-MFR creation'' formation scenario with magnetic field extrapolations. 

In fact, there have been some studies that used  NLFFF extrapolations to infer the role of ideal MHD instabilities of pre-eruptive structures. For instance, \citet{Guo_etal_2010}  found indications for Kink Instability in a confined eruption. Also, \citet{Inoue_etal_2011, Inoue_etal_2012, Inoue_etal_2013} analyzed the pre-eruptive configuration from NLFFF extrapolations by characterizing the twist of individual fieldlines numerically, based on the theoretical work of \citet{Berger_Prior_2006}. They found indications for the Torus Instability as the  trigger of eruptive events.

Here, we present unique SDO observations and magnetic field analysis that suggest the formation of MFRs at least 12 hours before they erupt as CMEs. In other words, confined flaring/sub-flaring builds up stressed magnetic fields, which eventually give rise to MFRs that erupt. \citet{Patsourakos_etal_2013} and \citet{Tziotziou_etal_2013} proposed this confined-flare-to-MFR formation scenario solely from observations without support from NLFFF extrapolations.

According to the confined-flare-to-MFR formation scenario, reconnection takes place first in a non-eruptive manner; this changes the topology, i.e. arcade to flux rope conversion, and heats the plasma within the flux rope structure to flare temperatures. This is why we can observe the hot flux ropes during confined flares.  Thus the flaring serves as a marker of topology change and illuminates the resulting magnetic structures. Flux cancellation as described in \citet{vanBallegooijen_Martens_1989} or \citet{Green_etal_2011} also involves magnetic reconnection and arcade-to-flux rope conversion, but requires the disappearance of magnetic elements in the photosphere; a process which is not always observed.

\begin{figure*}
\epsscale{2.0}
\plotone{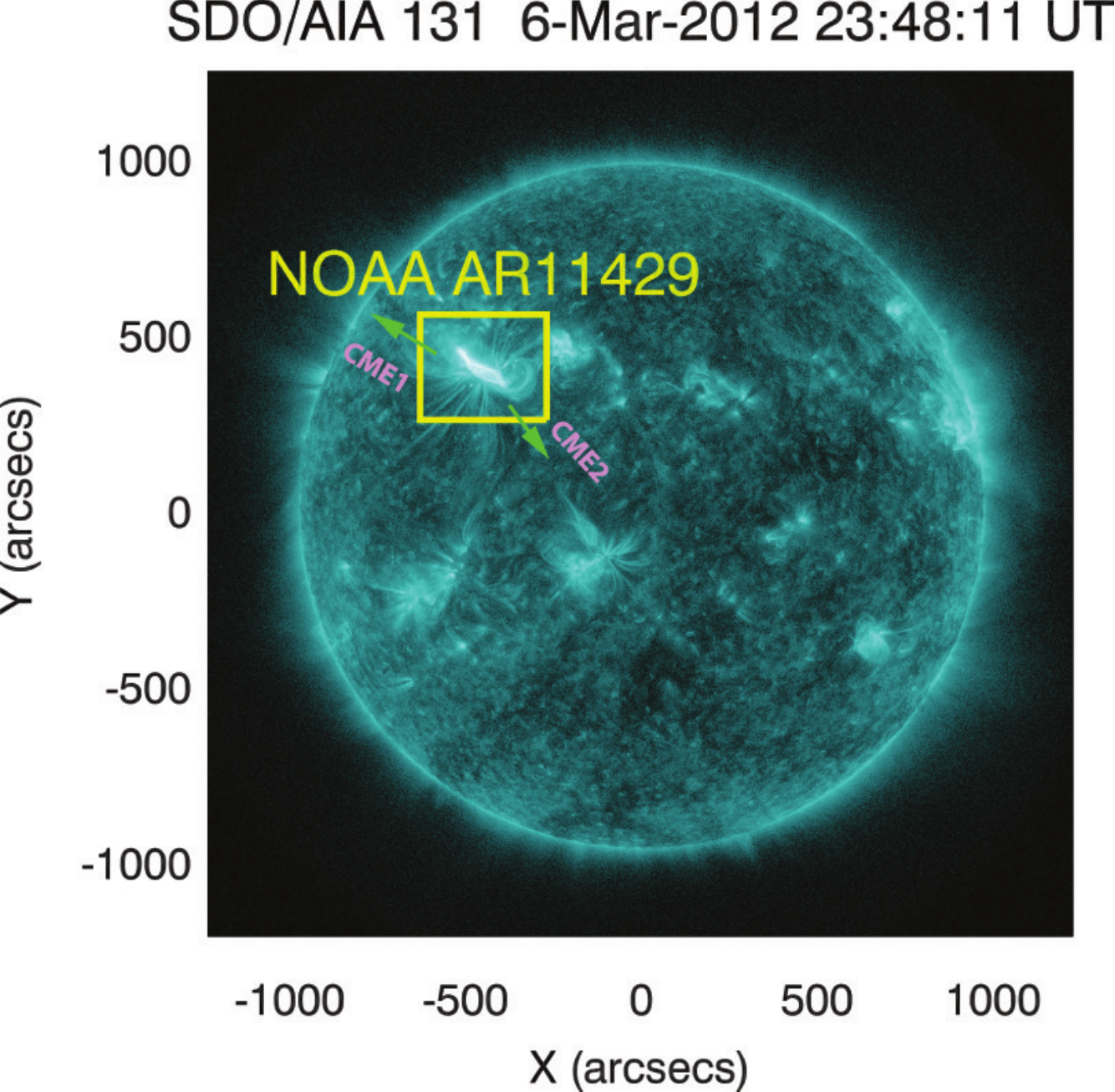} 
\caption{
Full disk image from SDO/AIA at 131\,\AA\ identifying AR11429. The size of the box is 500$\arcsec \times$400$\arcsec$. The direction of the CMEs at the onset of the eruptions is shown with the green arrows.
}\label{FIG_AIA}
\end{figure*}

The source region was $\delta$-spot NOAA AR11429  (for a full-disk AIA context image see Figure~\ref{FIG_AIA}). AR11429 was a ``super-active'' AR, which ejected several CMEs and a multitude of flares during its transit over the solar disk.  We focus on two fast CMEs ($\geq$ 2\,000km/s) which occurred within one hour from each other on March 7, 2012. These CMEs, which were initiated from different parts of AR11429 and followed different propagation paths in the low corona and beyond, are of general interest as they were associated with one of the most intense geomagnetic storms of cycle 24 to date ($D_{st}=-143$\,nT; \citealt{Richardson_2013}).

The  paper is structured as follows; in section 2 we provide an overview of the observations. In section 3 we discuss the coronal activities from the time of the confined flaring to the time of the twin CME eruptions. The NLFFF extrapolation specifics and the associated analysis and findings from magnetic data products are discussed in section 4. In section 5 we discuss the evidence supporting the ``dormant'' MFR scenario.

\section{Overview of Observations}\label{OBS} 

We use high temporal cadence and spatial resolution observations taken in different spectral bands by the Atmospheric Imaging Assembly (AIA; \citealt{Lemen_etal_2011}) and the Helioseismic and Magnetic Imager (HMI; \citealt{Scherrer_etal_2012})  aboard the \emph{Solar Dynamics Observatory} (\emph{SDO}; \citealt{Pesnell_etal_2012}). SDO orbits Earth at an inclined geosynchronous orbit. For three weeks around the Vernal/Autumnal equinox, the solar disk is briefly obscured by the Earth on a daily basis, in what is known as the \emph{SDO eclipse season}. Since the events under study were in early March this introduced daily data-gaps in our observations.
For a different vantage point, we used images from the Extreme Ultraviolet Imager (EUVI B for short; \citealt{Wuelser_etal_2004}) at $\lambda$ 195\,\AA\, part of the Sun-Earth Connection Coronal and Heliospheric Investigation instrument (SECCHI; \citealt{Howard_etal_2008} aboard the Solar Terrestrial Relations Observatory (STEREO; \citealt{Kaiser_etal_2008}). The STEREO mission comprises two satellites, one moving ahead (STEREO ``A'') and one behind (STEREO ``B'') the Earth. Only STEREO-B was appropriately positioned at that time. At the time of the observations STEREO B was at $\approx 117\degr$ east of the Sun-Earth line. 
Both AIA and EUVI-B as well as HMI are full-disk imagers of the coronal plasma and the photospheric magnetic field respectively.

\subsection{Magnetic Field Observations in the Photosphere}\label{HMI_OBS}
For the vector magnetic field observations we used preprocessed cutouts (containing the AR) at full-cadence (720 sec). The preprocessing consists of the resolution of the 180$\degr$-ambiguity for the azimuthal component of the magnetic field, and the transformation of the helioprojective images to a Cylindrical Equal Area (CEA) projection, which preserves the pixel size (\citealt{Hoeksema_etal_2014}). The CEA projection is essentially a transformation of the observational data into a Cartesian geometry, as if we were observing the AR ``from above'' while commoving with its guiding center. The resulting CEA dataset comprises three sets of 2-D Cartesian images, that is, one image for each component of the photospheric vector field, $\mathbf{B_{phot}}=$ (B$_x^{phot}$, B$_y^{phot}$, B$_z^{phot}$). The physical pixel size in the CEA dataset is 0.36\,Mm\,pixel$^{-1}$ and the size of the original cutouts was reduced to 205\,Mm$\times$145\,Mm (or 564$\times$391 pixels) which enveloped the AR. Due to the eclipse season, the total number of vector cutouts is reduced to 228 frames (instead of 240). A sample vector image from this time-series (showing all three components of $\mathbf{B_{phot}}$) is presented in Figure~\ref{FIG_HMI}.

Figure~\ref{FIG_HMI} is illustrating the complexity of AR11429. It is an \emph{anti-Hale} AR, which means that the leading polarity is opposite to that of the hemispheric trend. In fact AR11429 is the strongest (in terms of total unsigned magnetic flux; $\Phi=5.9\times10^{22}$\,Mx) anti-Hale AR of Cycle 24 to date.
It exhibits a complex, $\delta$-spot configuration (specifically, it is $\beta/\gamma/\delta$-configuration). $\delta$-spots are known for their extreme activity (e.g. \citealt{Zirin_Tanaka_1973}; \citealt{Tanaka_1991}; \citealt{Fan_2009} and references therein). 

Another interesting feature of AR 11429 is the existence of a complex,  sharply-defined and evolving PIL as seen in the $B_z$ component of the photospheric magnetic field. In particular, the western part of the AR exhibits multiple  PILs (Fig~\ref{FIG_HMI}, SW Box), in contrast to the eastern part of the AR which features a simpler and longer PIL. So we have the rare occurrence of PILs of different spatial scales coexisting within the same AR, and all of them exhibiting strong spatial gradients (i.e. very dense PILs). 

A useful metric for the strength of the spatial gradient associated with PILs is the R-parameter (\citealt{Schrijver_2007}). It parameterizes the unsigned magnetic flux near high gradient, strong field PILs from LOS magnetograms. In our case, we get a $log_{10}R \approx$ 5.2 which lies at the high-R tail of Schrijver's statistical study. However, the exact value of R might differ depending on the instrument (MDI vs HMI) and their properties (filling factor, resolving power and point spread function).  Nevertheless, AR 11429  has a strong gradient PIL which is quite often seen in $\delta$-spot ARs. Also, in Figure~\ref{FIG_HMI} one can see that the horizontal field vectors (shown with blue color) close to the PIL are oriented along the PIL. This is evidence of strong non-potential shear and we discuss this in more detail in Section \ref{SHEAR}. 

In Figure~\ref{FIG_EVOL} we present the time evolution of the photospheric vector field during March 6, 2012. 
It is quite obvious that significant changes occur along the PIL by means of shearing motions of individual flux elements (green circles; most notably in the NE side of the PIL - white ``blobs''). These ``blobs'' originate from a small emergence event on March 5, 2012, 10:30 UT close to the Northern negative polarity of the AR. In turn, the shape of the PIL (shown in red; found for every frame by means of an image gradient operation as in \citealt{Zhang_etal_2010}) is changing as this can easily be seen in Fig~\ref{FIG_EVOL}. Shearing also occurs at the location of the SW negative sunspot with the elongated positive polarity on top if it.

Another interesting trend of the photospheric evolution is the convergence of the southmost negative sunspot towards the PIL. In addition to all these motions, AR 11429 displays sunspot rotation, which cannot be easily appreciated in Fig~\ref{FIG_EVOL}. To better appreciate the reported PIL dynamics, a movie of CEA vector frames spanning a period of 5 days is available in the online version of the paper (filename ``movie1.mpg''). 
To summarize, these observations suggest that the shearing motions and sunspot rotation contribute to the build-up of the non-potential shear angle in the photosphere, and of stressed magnetic field configurations in the corona. The importance of these observations will become apparent in Section~4. 
\subsection{Magnetic Field Extrapolation}\label{EXTRAPOL}

Due to the on-disk location of AR 11429, it is possible to use photospheric magnetic field extrapolations to assist our investigation on the structure of the coronal magnetic field. A reasonable approximation for the magnetic fields in the corona is the Force-Free assumption, i.e. the absence of Lorentz Forces by neglecting the cross-field electric currents (for a discussion see review by \citealt{Wiegelmann_Sakurai_2012}). In other words, the model requires that the electrical currents flow strictly along the magnetic field lines. Another requirement 
is the absence of magnetic monopoles in the bounded volume, V. Using vector notation, the above can be expressed as

\begin{equation}
\nabla\times\mathbf{B} =\alpha(x,y)\,\mathbf{B}\ ; \ \  \nabla\cdot\mathbf{B} =0
\end{equation}
\\
where $\alpha(x,y)$ is the force-free or torsion parameter, which in general is a function of position but is conserved along each field line.  The above set of equations is describing the general case, i.e. that of Non-Linear Force-Free Fields (NLFFF), which is considered quite successful in reproducing the overall morphology of ARs and current channels along PILs (\citealt{Wiegelmann_Sakurai_2012} and references therein). A special case is the linear force-free case, when $\alpha(x,y)=const$ globally, which is known to be inaccurate when the extrapolated magnetic field lines are compared to structures from EUV images (\citealt{Wiegelmann_Sakurai_2012}).

For our investigations we used an optimization technique (see \citealt{Wiegelmann_2004}) for computing the NLFFF in the corona, based on the method introduced by \citet{Wheatland_etal_2000}. Using appropriate boundary conditions (see following subsection \ref{inp_hmi} and Appendix A), this numerical method yields a NLFFF solution by minimizing a penalty function, $L$, in the computational volume, V, as

\begin{equation}
L= \int \limits_V \, w(x,y,z)[B^{-2}|(\nabla\times \mathbf{B})\times \mathbf{B}|^{2}+|\nabla \cdot \mathbf{B}|^{2}]\, \mathrm{d} V
\end{equation}
\\
where $w(x,y,z)$ is a scalar function with a value of 1 in the physical domain of the volume and drops smoothly to zero when approaching the top and lateral boundaries. Obviously, when the penalty function $L=0$, both the Lorentz force is zero and the solenoidal condition is satisfied in the entire computational volume, V, which now contains the NLFFF. 

\subsubsection{Input Data to the NLFFF Code}\label{inp_hmi}
The input data for the NLFFF calculation were photospheric vector magnetogram cutouts from the HMI (discussed in section~\ref{OBS}). We performed NLFFF extrapolations at an hourly time-step, covering the period from March 6, 2012 00:00 UT to March 7, 2012 23:00 UT. Due to the eclipse season, extrapolations at 08:00 UT of both March 6, 2012 and March 7, 2012 were not calculated. Moreover, the boundary of March 7, 2012 at 07:00 UT was determined to be problematic (out-of-focus) and it was  discarded. We ended up with 45 NLFFF extrapolations (3-D vector datacubes) per time-step. The size of the NLFFF extrapolation datacubes used in our analysis was 250$\times$163$\times$106 pixels (or 180$\times$117$\times$76\,Mm$^{3}$) for each component of the NLFF magnetic field (see appendix for more information).

\subsection{Coronal Observations}
We analyze the AIA coronal observations from March 5, 2012, 23:59:58 UT, when the AR was at E43$\degr$N18$\degr$, to March 6, 2012, 23:59:58 UT E17$\degr$N18$\degr$. The spatial resolution of the AIA images is $0\,\farcs6$ pixel$^{-1}$. The passbands used (and their nominal formation temperature) are: 131\,\AA\ (0.4\,MK \& 10\,MK), 94\,\AA\ (6\,MK), 211\,\AA\ (2.0 MK), 193\,\AA\ (1.6\,MK \& 20\,MK). The time-cadence is 12 sec, resulting in 14400 images per passband. Two data gaps exist (one per day between 06:50 and 07:40~UT) due to the SDO eclipse season. From the full-disk images we extracted a sub-field centered at the AR at a size of 600$\arcsec\times$500$\arcsec$. The Field of View (FOV) size was chosen to fully contain the AR, while it rotates at the Carrington rotation rate. 

From the four available passbands of EUVI-B, we selected the 195\,\AA\ passband because it has the highest cadence (5 min) and probes the hottest plasma. The 195\AA\ channel contains a weak Fe XXIII line which contributes only during flares and is of interest to us since we are searching for a hot MFR. The pixel size is 1$\farcs\,$59. There were 576 images in total and the AR was located close to  the EUVI-B western limb. The combination of SDO/AIA and EUVI-B observations provided a comprehensive view of AR 11429 from above and from the side, respectively. 

\section{Coronal Activity Prior to the CMEs}\label{CORONAL}

As we mentioned before, the transit of AR 11429 is marked by intense coronal activity with many eruptive and confined flare events. During the period of interest here, covering the entire day before the onset of the CMEs of March 7, 2012, the GOES satellite records several SXR flares above C level in the 1-8\,\AA\ X-ray flux (Figure~\ref{FIG_GOES}). 
We use the 131\AA\ images, covering AR 11429, to locate the source of this activity. It  manifests itself as transient brightenings, either strong enough to be associated with a GOES flare or weaker in the sub-flare domain, of hot ($\sim$10\,MK) coronal structures. The close association between the 131\AA\, light-curve of AR 11429 and the GOES light-curve suggest that the SXR emissions originate from this AR.

The brightenings occur predominately in two locations, in close proximity to the PILs and have a very elongated character (Figure~\ref{FIG_AIA_EVOL}). To examine their relationship to the underlying complex magnetic field structure, we formed an AIA/HMI composite movie (filename ``movie2.mpg'') using the 131\,\AA\ and the Line of Sight Magnetogram (BLOS) images (blue negative, orange positive). We show a set of representative snapshots in Figure~\ref{FIG_COMPOSITE_MULTI} and the movie is available in the online version of the paper. The BLOS background images are saturated at $\pm30\,G$ in order to serve as a PIL-tracer, which is found at the interface between the two color-coded polarities; the foreground 131\,\AA\ image shows that the transient brightenings reside in the intersection of the blue and orange saturated areas of BLOS, i.e. in the vicinity of the PIL. The movie spans five days and a large range of heliographic longitudes. The close proximity of the brightenings with the PIL, over such a long time,  is indicative of low-lying active structures. 
Thus, the 131\,\AA\ evolution indicates two main kernels of activity within the AR. One kernel is associated with the NE PIL and the other with the SW PIL. In fact, the SW kernel is active all the time -i.e. bright in the 131\,\AA\ passband-, while the NE is active but in a more intermittent fashion. In the beginning of March 6, 2012 the NE brightenings seem to originate from a fragmented structure, apparently small hot-loop-channels tilted with respect to the NE PIL. 
The structures traced by the transient brightenings become more monolithic---they transition into an elongated, possibly low-lying structure, almost perfectly aligned with the NE PIL, towards the end of March 6, 2012. Another important characteristic of the hot-loop-channels seen in Figures~\ref{FIG_AIA_EVOL} and~\ref{FIG_COMPOSITE_MULTI}, is that the highlighted structures (either pointed with an arrow or dotted lines) have a weak sense of an S-shape and hence maybe weakly twisted. 

\subsection{Confined Flaring Activity during March 6, 2012}
12 hours prior to the CMEs of March 7, 2012, an M2.1 flare occurred around 13~UT on March 6, 2012 (labeled ``confined'' in Figure~\ref{FIG_GOES}).  As discussed earlier, the transient activity, in the form of brightenings in 131\AA, is continuous and localized at the NE and SW PIL (Figures~\ref{FIG_AIA_EVOL}-\ref{FIG_COMPOSITE_MULTI}). The M2.1 flare brightens a large part of the NE PIL and the corresponding structure seen in 131\,\AA\ (e.g. top right panels of Fig~\ref{FIG_AIA_EVOL}, dotted line; Fig~\ref{FIG_COMPOSITE_MULTI}, red arrow) gives the impression that is being lifted up. It becomes fainter as it (presumably) cools. The same reappears shortly before the onset of the first CME of March 7, 2012 and eventually erupts with this CME. The location of M2.1 flare lies in the middle of the NE PIL, as determined from the 131\,\AA\ image on March 6, 2012, 12:35 UT and is the same location for the eruptive X5.4 flare at 00:02 UT of March 7, 2012 associated with our first CME. 

There were no large-scale EUV dimmings associated with the M2.1 flare as can be seen in 211\,\AA\ base difference images in Figure~\ref{FIG_BDIFF}. A similar result was reached with 193\AA\ base difference images. No CME was recorded in the available coronagraph data (SECCHI, LASCO). The lack of large-scale EUV dimmings and coronagraph CME signatures lead us to conclude that the M2.1 flare  was a confined flare. 

One hour after the  peak of the confined M2.1 flare, a Differential Emission Measure (DEM;  following \citealt{Plowman_etal_2013}) analysis verifies the existence of a hot ($\sim$10\,MK) structure in the NE PIL (bottom 6 panels of Figure~\ref{FIG_DEM}).  The structure is also seen in the 131\,\AA\  images taken after 12:23~UT (Figure~\ref{FIG_AIA_EVOL}). We interpret this as the formation and filling with hot plasma of an MFR structure during the M2.1 flare. The overall shape and morphology of the NE structure associated with CME1 is very similar to the one seen at 12:23~UT. This is demonstrated in the top right and bottom right panels Figure~\ref{FIG_AIA_EVOL} (dotted yellow lines are added to envelope the bright structure).

Detailed spectroscopic diagnostics of AR11429 by the EIS spectrometer on Hinode, around 12:00 UT and 21:00 UT on March 6, 2012 also show evidence for hot MFR structures (Syntelis, Gontikakis, Patsourakos, Tsinganos 2015; in preparation). 

A second M-class flare (M1.3) occurred around 21~UT on March 6 (Figure~\ref{FIG_GOES}) at the same location as the first M-class flare. For the sake of brevity, we do not describe the analysis details here but the coronal response was very similar to the first M-class flare -- i.e, no CME, a confined flare and ``illumination'' of MFR-like structures. We thus have two confined M-class flares in the midst of a plethora of transient brightenings resulting in the creation or illumination of a hot MFR structure at the location of a fast CME 12 hours later. 

Regarding the SW PIL, there is persistent sub-flaring activity with small transient brightenings which are more spatially constrained than those in the NE PIL (due to its smaller lengthscales). There were no CMEs originating from that location during March 6, suggesting that the observed sub-flarings were confined. The activity may indicate the sites of stressed magnetic structures that could have contributed in the formation of the MFR of CME2 (see Fig~\ref{FIG_COMPOSITE_MULTI}). 
This PIL is the location of the second X flare on March 7, $\sim$ 50 mins after the X flare at the NE PIL.


Observations from the Earth (SDO/AIA) and from a different viewpoint (from the 195\AA\ channel of EUVIB) suggest the slow rise of the system during the M-class flare of March 6, 2012. To measure the speed of the rise we created slit-time plots seen in Fig~\ref{FIG_EUVIB}. The position and orientation of the slit was selected to intercept the images along the direction of fastest expanding motions (in the online version of the paper a multipanel movie is available, ``movie3.mpg'').
The measured speeds of the bright front seen in the right panels of Fig~\ref{FIG_EUVIB} are $\sim$ 10-20km/sec (derived from the 
slope of the ``step''-like feature in the slit-stack maps covering 11:00 - 15:00 UT of March 6, 2012; left panels). 
These speeds represent a small fraction of the Alfv\'{e}n speed in the corona, 
and signify a slow quasi-static rise of a magnetic structure possibly caused by slow photospheric motions (shearing or twist). We essentially have a two-phase kinematic behavior around the flare event.  First, a quasi-static slow rise (the ``step''-like feature in the slit-stack maps) at 10-20 km/s from $\sim$ 12:23~UT during the flare.  The rise phase ends at 12:40~UT. Thereafter we observe much shallower slopes signifying that the magnetic structure has reached a stable position. This is additional evidence for a confined flare. 


In the cooling phase of the M2.1 flare ($\sim$13:46), cool filamentary plasma appears along the presumed MFR axis. The red arrows in Figure~\ref{FIG_EUVI_ABSORPT} show a dark absorption feature and hence cooler than the peak temperature of 195\,\AA, 1.6MK. This suggests that a coherent structure exists along the PIL to support cool plasma. This observation is in agreement with MFR observations of \citet{Patsourakos_etal_2013} that show cooling at the MFR center. Note that the same absorption can be seen from the AIA vantage point (almost seen from above, top right panel in Figure~\ref{FIG_COMPOSITE_MULTI}).

\section{MFR Tracers from the NLFFF Extrapolations}

To further aid in our search for pre-existing MFRs, we turn to the NLFFF extrapolations. We use them as a ``laboratory'' to generate several diagnostics to explore the existence of MFRs in the pre-eruptive magnetic configuration. In the following paragraphs we discuss these diagnostics separately.

\subsection{Magnetic Stream-line and Electric Current Visualization}\label{stream_line_section}
In Fig~\ref{FIG_EXTRAPOL} we show the extrapolated NLFFF by means of magnetic stream lines on March 6, 2012, 23:48 UT as seen from the top of the domain. The time corresponds to shortly before the onset of the first CME. For the purpose of clarity and in order to give a sense of 3-D depth, we color the magnetic stream lines with green color for the overlying field and with teal color for the lower lying magnetic structures. In the online version of the paper a movie showing a fly-by around that extrapolation is available (``movie4.avi''). The low-lying magnetic structures exhibit MFR morphology when observed very close to the PIL and they gradually become sheared for lines that are tied further about the PIL. The overarching, higher lying magnetic structure seems to describe well the EUV images and seems closer to a potential state. As suggested from the 131\AA\ brightenings, the majority of the transient activity originates from low-lying magnetic structures along the PIL. Strictly speaking, there are two main kernels of activity, one in the NE PIL and another in the SW PIL (as seen in the previous figures). 

By applying Amp\`{e}re's law as $\nabla\times\mathbf{B_{NLFFF}}=\mu_0 \mathbf{J}$ numerically, we get the spatial distribution of electrical currents $\mathbf{J}$ in the NLFFF. Their magnitude, $|\mathbf{J}|$, represents a proxy for ohmic heating. In other words, regions with strong $|\mathbf{J}|$ would be more heated and thus exhibit higher temperatures. A more complete approach should incorporate a full energy equation, which includes source (e.g., Ohmic heating) and cooling (e.g. radiation and conduction) terms to work out the detailed plasma emission in any AIA channel. However, a careful inspection of the 131\,\AA\ image time series with images containing the areas of strong electrical currents in the NLFFF datacubes shows that the 10\,MK plasma emission is described well by the location of the stronger NLFFF-inferred electrical currents. The strength of these currents is characteristic of the compactness of the PIL. 

We created $|\mathbf{J}|$ time series which contain the LOS integral of $|\mathbf{J}|$ (middle row panel of Fig~\ref{FIG_MULTI_J}). Also we took vertical planar cuts in the middle of the NLFFF boxes that show $|\mathbf{J}|$ in color contours with the local magnetic field projected onto the cut plane (bottom row panel of Fig~\ref{FIG_MULTI_J}; the leftmost ``lobe'' corresponds to the NE MFR and the rightmost to the SW MFR). The oval shape of the B-projection suggests the existence of MFR structures detached from the photosphere. The cores of the structures are low-lying ($\approx$ 3\,Mm). The length of the two prevalent current channels is 80\,Mm for the NE MFR structure and 40\,Mm for the SW. The radius of the MFRs is about 10\,Mm (NE) and 6\,Mm (SW) (deduced from where $|\mathbf{J}|$ is 20\% of the peak current). The mean total magnetic field strength measured on the cut plane of the NE MFR (before the eruption) is $\sim$ 650\,G and for the SW MFR, $\sim$ 500\,G (see Table~\ref{TABLE}).


There is also a striking similarity between the 131\,\AA\ and $|\mathbf{J}|$ time series of Fig~\ref{FIG_MULTI_J}. At the beginning of our time series (early March 6, 2012) the transient brightenings along the NE PIL  delineate a fragmented active structure. As the time goes by and towards late March 6, the NE PIL is undergoing transient brightenings, but now as a more monolithic structure. This is evidence for highly-sheared (almost parallel to the PIL) structures that seem to comprise a very elongated structure along the PIL. This could be a result of a continuous cancellation process which creates a very elongated structure with a strong axial field (\citealt{vanBallegooijen_Martens_1989, Green_etal_2011}). This fragmented-to-monolithic current channel formation is reflected in the NLFFF $|\mathbf{J}|$ datacubes. 

Persistent transient brightenings also occur in the shorter, SW PIL. There is a very good correspondence of the location of NLFF volumetric currents with the location of the brightenings in 131\,\AA. However the picture is more complicated when inspecting the vertical cuts; there seems to be a low lying persistent structure, or ``island'', between the structures associated with the NE and SW MFR. This structure exhibits an opposite sense of rotation, or chirality, for the magnetic field lines. More details can be found in Section~\ref{chiral}.

The overall picture we get from the location of the two strong current channels along the PIL and the helical morphology of the magnetic field around these current channels is suggestive of the existence of two MFR candidates.
 

\subsection{Twist Number and Mean Twist Evolution} \label{twist_section}
An alternative way to detect twisted magnetic field structures 
in NLFFF extrapolations was introduced by \citet{Inoue_etal_2011, Inoue_etal_2012, Inoue_etal_2013}. 
This numerical technique is based on the theoretical framework of \citet{Berger_Prior_2006} 
in which they derived a simple expression for the twist number, $T_n$, for magnetic field lines with non-zero helicity, as
\begin{equation}
T_n=\frac{1}{4\pi}\int \alpha\,\mathrm{d}l 
\end{equation}
\\
where, $\alpha$, is the force-free parameter, and, $dl$, is the infinitesimal length of the field line. For force-free fields, $\alpha$ is constant along a field line and the twist is then given by 

\begin{equation}
T_n=\frac{1}{4\pi} \alpha L
\end{equation}
\\
where now, $L$, is the total length along a field line. We derive $\alpha$ as $\nabla \times \mathbf{B}|_{phot}$. Thus, we can obtain the twist number for each field line since we know its length $\alpha$ from NLFFF extrapolation.

The method by \citet{Inoue_etal_2011} is capable of producing 2D maps of the twist number, $T_n$,  by mapping $T_n$ at the footpoints of each magnetic line by simply multiplying $\alpha$-maps and ``length-maps''. However, caution must be taken since $\alpha$ may not be the same at both footpoints. This can be either due to instrumental noise, deviations from force-free condition or noise due to numerical derivatives associated with the calculation of $\alpha$. In our  implementation, we simply take the average value of the $\alpha$ between the negative and positive polarities as $\bar{\alpha}=\frac{\alpha_{+}+\alpha_{-}}{2}$ and then we smooth the map with a 3$\times$3 gaussian kernel (following  \citealt{Inoue_etal_2011}). For the line integration we use a second order Runge-Kutta integrator with a $1/16$-pixel integration ``time-step''. The line integration starts from pixels below a threshold $B_z^{thresh} \leq-$500\,G (i.e. from negative footpoints).  When the line integration reaches the positive polarity at the photosphere, the integration stops and the value of total length of the field line is stored on a 2-D map on both positive and negative footpoints. We call this the ``length-map''. After the entire negative polarity ``seedpoints'' have been used for the integration, the map looks smooth in the negative polarity but the positive polarity seems ``sparse'' (i.e. contains holes, in other words, points of avoidance for the final locations of the integration). This signifies that the NLFFF method does not perfectly satisfy the solenoidal condition at the boundary, thus we resorted to a work-around described in the appendix B. 

A major concern using $T_n$ maps for detecting MFR structures is the possibility of confusing Sheared Magentic Arcades (SMAs) with MFRs. So we need to define $T_n$ thresholds separating MFRs and SMAs. For this purpose, we used an analytical expression (\citealt{Aschwanden_2004}) to derive various SMA fields and found that they could reach a maximum $|T_n|$ of 0.5. So we considered as an MFR field line any line with $|T_n|\geq$ 0.5. 
Our choice of the absolute twist threshold between a SMA and an MFR seems reasonable and conservative. 
The inspection of the $T_n$ maps ( Fig~\ref{FIG_TWIST_MAP}) reveals that: (1) we obtain a few locations of relatively high twist around and along the PIL in accordance with the imaging observations in the corona, and  (2) $|T_n|_{max}$ is not extremely big (not significantly exceeding one full turn). These findings are indicative of weakly twisted low-lying MFRs, in agreement with the EUV observations in the hot 131\,\AA\, channel.

In Fig~\ref{FIG_TWIST} we show the temporal evolution of the spatial averages of the force-free parameter, twist number, and the fraction of twisted flux above selected twist thresholds over total flux, defined as 

\begin{equation}
F(\tau)=\frac{\int_{|T_n|>\tau} B_z \mathrm{d}S}{\int B_z \mathrm{d}S}
\end{equation}

\noindent The surface integral is taken over the negative polarities. 

The analysis is showing an overall increasing trend for the average $\alpha$, $T_n$ and F($\tau$) over the  entire considered time period. For the average force-free $\alpha$ there seems to be a decreasing trend before the onset of the CME events (vertical line at 00:00 UT) and a rapid increase after the events. The average $T_n$ shows a steady value of about a quarter turn with negative chirality which increases slowly after the CMEs. The fractional twisted flux F($\tau$) for $|T_n|\geq$0.5 (half a turn) shows qualitatively the same behavior to the average twist. However the fraction of twisted flux above half a turn is 10\% and steadily increases after the CMEs. The interpretation for this is due the continuous growth of the AR in terms of magnetic flux (in the form of emergence of small magnetic features throughout the time-range of the observations). Also,in addition to the continuous emergence,  the emerged flux is (or becomes) more stressed so that to be reflected in magnetic field metrics associated with its stress (e.g., twist number and alpha parameter).

Finally, in each of the top two panels of Fig~\ref{FIG_TWIST} we show the total number of footpoints in the negative seed-``patch'' (which were initially selected as $B_z\leq-$500 G). There is an increase of the number of points (from 8700 to 9500 points) after the CMEs which implies that there was an increase in the magnetic flux at the photospheric boundary which supports our interpretations of the previous paragraph.

\subsection{Photospheric and Coronal Chirality Patterns} \label{chiral}

Since the structures undergoing transient brightenings in 131\,\AA\ (Figures~\ref{FIG_AIA_EVOL} and \ref{FIG_MULTI_J}) are strongly emitting and stand out well about the background, it is possible to identify the helicity sign, or the \emph{chirality} of the structures and compare it with the NLFF results. The chirality can be obtained, as a 2-D map, from the sign of the force-free parameter, $\alpha$. For this task, we convert the chirality maps from the native heliographic (CEA) coordinates of the NLFFF into helioprojective. Now, we can identify where the ``active'' (in 131\,\AA\,) PIL-aligned structures are rooted and thus estimate their chirality. Since these are transient brightenings, their intermittent nature can be summarized best in a single map by calculating the standard deviation of the 131\,\AA\ time-series for each pixel.

In Figure~\ref{FIG_ALPHA} we present composites of $\alpha$-maps and 131\,\AA\ standard deviation maps accumulating activity for two hours before the confined M-class flare and the CME eruptions. 
Again, the maps show two clear clusters of high-temperature coronal activity: one in the NE PIL and the other in the SW PIL. The dominant chirality of the NE PIL ``structure'' or cluster in the 131\,\AA\ is negative (overlaid on red $\alpha$-patches) but for the SW PIL the picture is complicated. In fact, there is a patch or ``island'' of positive chirality  (blue $\alpha$-patch) in between the negative chirality of the NE PIL and the negative chirality at the SW PIL (see also vertical cuts at bottom panels of Fig~\ref{FIG_MULTI_J}; the positive chirality structure corresponds to the middle current structure). 

In addition, Figure~\ref{FIG_ALPHA} shows that this is the location of continuous activity of small low-lying structures. This holds true throughout our observing period. Transient brightenings indicate continuous, small-scale reconnection events (no big flares during the considered intervals) localized within the two distinct regions---one in the NE and the other in the SW. Thus, the chirality patterns support the existence of 2 MFRs in the location where the two CMEs occur.

\subsubsection{Shear Angle Evolution}\label{SHEAR}

Another way to characterize the deviations from potentiality is by means of the shear angle of the photospheric magnetic field on the PIL (\citealt{Hagyard_etal_1984}; \citealt{Ambastha_etal_1993}). The shear angle is defined as the angle between the observed photospheric field and the potential field at the photospheric layer,

\begin{equation}
\theta=cos^{-1}\left(\frac{\mathbf{B}_{obs}\cdot \mathbf{B}_{pot}}{|\mathbf{B}_{obs}||\mathbf{B}_{pot}|}\right)\bigg|_{phot} \label{shear_eq} 
\end{equation}

The larger the shear angle, the further from the potential state the magnetic field is at the PIL. The main process responsible for gradual shear build up is the slow photospheric motion of magnetic elements, which continuously drag (and shear) magnetic field lines. However, there have been reports for rapid, permanent changes of the shear during flares. Their interpretation is unclear as both a rise and a drop of the shear angle have been observed after the flare (\citealt{Wang_etal_1994}; \citealt{Petrie_2012}). The horizontal field in the photosphere could become more sheared following the formation of a vertical current sheet when a MFR moves upward (e.g. \citealt{Forbes_Priest_1995}). 
In our case, we calculated the shear angle $\theta$ using the eq~(\ref{shear_eq}) by sampling the shear angle along the PIL. 

The PIL is found for every frame by means of an image gradient operation (as in \citealt{Zhang_etal_2010}) and the end product is a binary map. The resulting PIL maps are grown to a thickness of 4 pixels. A histogram of the shear angle distribution versus time is shown in Fig~\ref{FIG_SHEAR}. Overall we get high shear angles and that is very important for the evolution of the AR, since shear can generate twisted field lines.

In the beginning of March 6, the mean shear angle is around 68$\degr$ and begins to slowly increase for about 6 hours before the CME events, reaching a maximum of 74$\degr$ at the moment of the eruption (left panel of Fig~\ref{FIG_SHEAR}). This number differs from the maximum number reported by \citet{Petrie_2012} ($\sim$ 90$\degr$) as they focused on a small part of the NE PIL. In the right panel, we show the total number of pixels above a threshold of shear angles. It is interesting that the number of pixels in the PIL with shear angles above 75$\degr$ is increasing steadily until the time of the CMEs; this is understood as there are more magnetic vectors in the PIL orienting along the PIL since if the structure was potential, the magnetic vectors would be vertically oriented to the PIL. These measurements indicate the presence of magnetically stressed structures in the locations of the two CMEs, which further suggests the existence of 2 parent MFRs.

\subsubsection{Critical Decay Index}

From the AIA observations at the time of the CME eruptions we deduce the initial directions of the CMEs in the low corona (directions shown in Fig~\ref{FIG_AIA}). To understand why the initial directions of the two CMEs differ although they originate from the same AR, we calculate the decay index, i.e. the decay rate of the strength of the horizontal magnetic field with height for the entire computational domain, defined as

\begin{equation}
n=-\frac{\mathrm{d} logB_h}{\mathrm{d} logz}
\end{equation}

where $B_h = \sqrt{B_x^2+B_y^2}$ and z is the radial distance measured outwards from the solar surface in solar radii. The decay index is a measure of how the horizontal field strength of the magnetic field decays with increasing height. This quantity has been used to define a threshold for the triggering of ideal-MHD instabilities, i.e. the Kink (\citealt{Torok_Kliem_2005}) and Torus (\citealt{Kliem_Torok_2006}) instabilities. For instance, for bipolar configurations and for $n \geq$\,1.5, an MFR is Torus-unstable  and thus can not be constrained by the overlying field and could erupt. In addition, the kink instability requires significant twist to be stored in the pre-eruptive magnetic configuration, $T_n\sim1.6$ (\citealt{Torok_Kliem_2005}).

In Fig~\ref{FIG_DECAY_INDEX} we show the distribution of the critical decay index at different heights before and after the time of the CME eruptions. For the decay-index calculations we used the NLFFF extrapolations field discussed in the previous paragraphs. The locations at each height (shown with different color contours corresponding to different heights) form closed contours and the big concentric ``holes''  denote the ``weak spots'' of the horizontal field, where the decay index is larger than the critical (or ``super-critical''; here $n \geq$\,1.5). This corresponds to areas with extremely weak horizontal component of the magnetic field (thus almost radial field lines), which is typically the case above the sunspots, e.g. at the leading sunspot in the west side of the AR. 

However, the magnetic configuration may allow for similar 'holes' away from the radial sunspot fields. For example, this is evident at the NE PIL (see family of black to blue contours), where a ``cone'' with decay index greater than the critical decay index is in close proximity to the MFR magnetic structure along the PIL. Thus, a structure which is becoming Torus-unstable would preferentially erupt, and most probably ``escape'' through this super-critical decay-index ``tunnel''. This can be seen in the three frames provided in Figure~\ref{FIG_DECAY_INDEX}. A movie covering the entire 2-day period of March 6-7 is available in the online version of the paper (``movie5.mpg''). A comparison of the panels focused on the NE PIL shows a migration and development of the supercritical decay-index ``tunnel'' stretched along the NE PIL (second panel). A comparison between the starting height of the ``escape'' tunnel (which before the eruption reaches all the way down to the surface), the cross-sectional radius of the NE MFR from NLFFF ($\sim$ 10\,Mm) and the apparent initial height of the erupting structure from EUVI-B (Fig~\ref{FIG_EUVI_ABSORPT}; $\sim$ 30\,Mm) yields consistent results supporting the necessary ingredients for an ideal instability to kick in. In addition, the site of the initiation
of CME2 corresponds also to low-lying supercritical decay-index regions.
A comparison between Fig~\ref{FIG_DECAY_INDEX} and the EUV observations shows good correspondence between the observed direction for the CMEs, and the locations of both the supercritical decay indices and our presumed MFRs. The critical index analysis, therefore, justifies the discrepancy in the two CME directions and provides additional, indirect support for the existence of MFRs since such structures are prone to plasma instabilities dependent on the critical index.

\section{Discussion and Conclusion} 
 
 We embarked on this study aiming to uncover the reasons behind this unusual twin CME occurrence.  The double CME events of March 7, 2012 originated from a very dynamic and rapidly evolving AR. Both CMEs were super-fast ($v_{CME} \geq$ 2000 km/sec),  launched within one hour of each other, and were associated with two major solar flares (X5.4 and X1.3). They, however, followed very different propagation paths. The Earth-directed CME, CME2, misidentified in previous works, caused the second larger geomagnetic storm of Cycle 24. The source AR exhibited all the hallmarks of a ``geomagnetically dangerous''  AR: it had an anti-Hale $\delta$-spot magnetic configuration, multiple and compact PILs, strong shearing motions and non-potential shear. In other words, AR 11429 had plenty of magnetic energy and complexity to expel multiple, fast CMEs, as it clearly did during its disk passage. 
 
 The continual flaring/CME activity of the source AR did not hinder our analysis, thanks to the detailed coronal coverage from SDO and STEREO. We were able to trace the origins of the double CMEs of March 7 to confined flaring events
of various magnitudes on the previous day. For example, an M2.1 flare on March 6 at 12:35~UT illuminated MFR-like structures to a 10 MK temperature (in the AIA 131 \AA\ images).  The position of the AR on the solar disk on March 6-7 (E30$\degr$N18$\degr$) posed some challenge in our analysis. The AR structures were viewed from above. As discussed in the Introduction, the background emission (from low-lying structures) can impede the identification of MFR structures in such projections. We focused on detailed NLFFF extrapolations using carefully treated HMI magnetograms over the full two-day period of interest. The extrapolations supported the presence of MFRs on March 6 as they showed MFR-like field lines---$|\mathbf{J}|$-concentrations corresponding to oval-shaped field lines in vertical plane cuts (Figure~\ref{FIG_MULTI_J})---  which are effectively weakly twisted structures with continuous twist build up.
 
\begin{figure*}
\epsscale{2.0}
\plotone{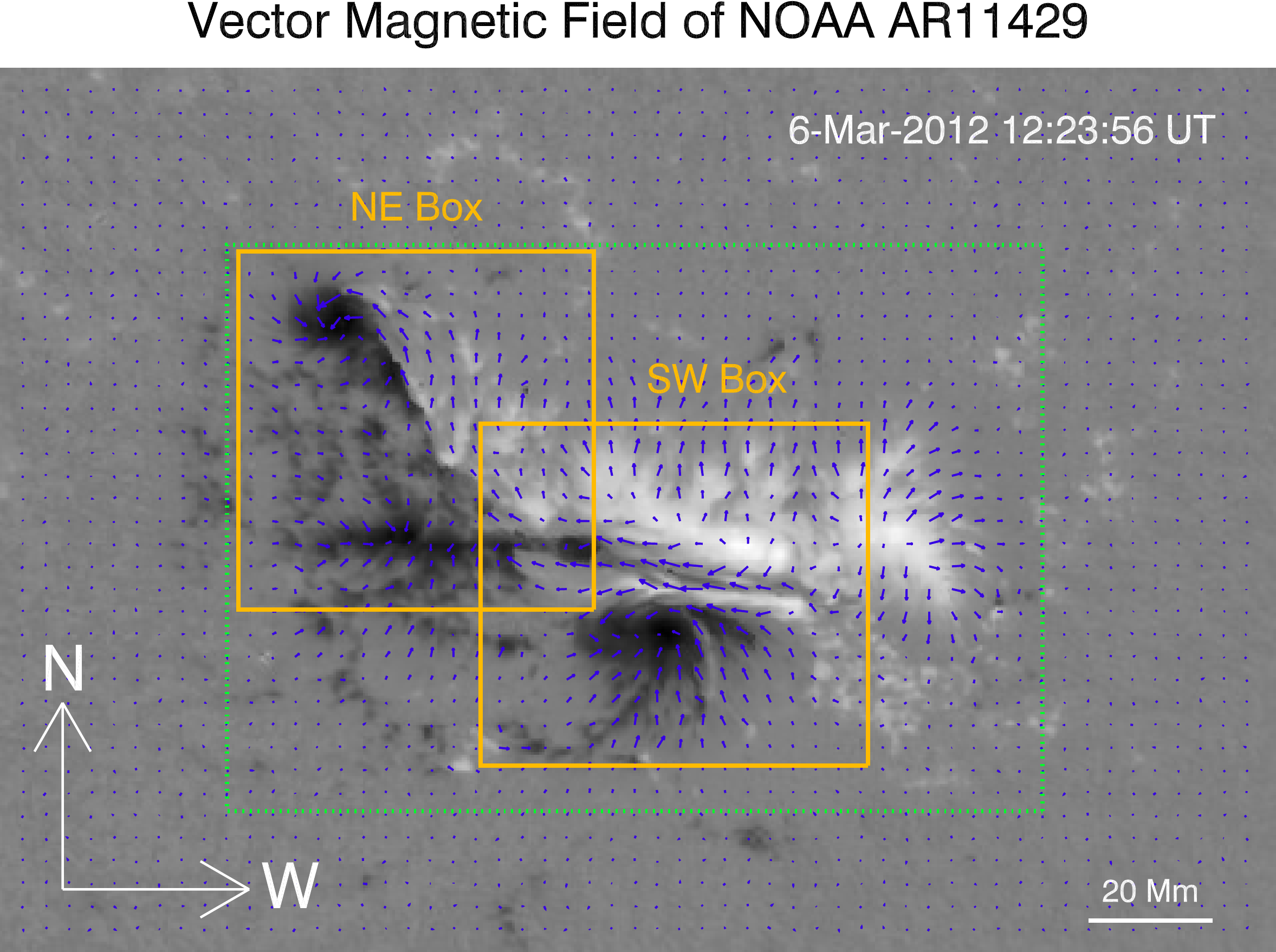}
\caption{
A sample CEA magnetic vector map from the HMI/SDO taken at the time indicated. The greyscale image is the normal component of $\mathbf{B}_{phot}$ saturated at $\pm$2500\,G. The horizontal photospheric magnetic field, $\mathbf{B}_h$ is shown with blue vectors. Note the alignment of the horizontal field along the PIL (observational manifestation of high PIL-shear). The two orange boxes shown enclose the NE and SW PIL respectively (boxes show FOVs used in Fig~\ref{FIG_TWIST_MAP}). The green dotted window was used for the calculation of the magnetic flux (in Fig~\ref{FIG_EVOL}).
}\label{FIG_HMI}
\end{figure*}

\begin{figure*}
\epsscale{2.0}
\plotone{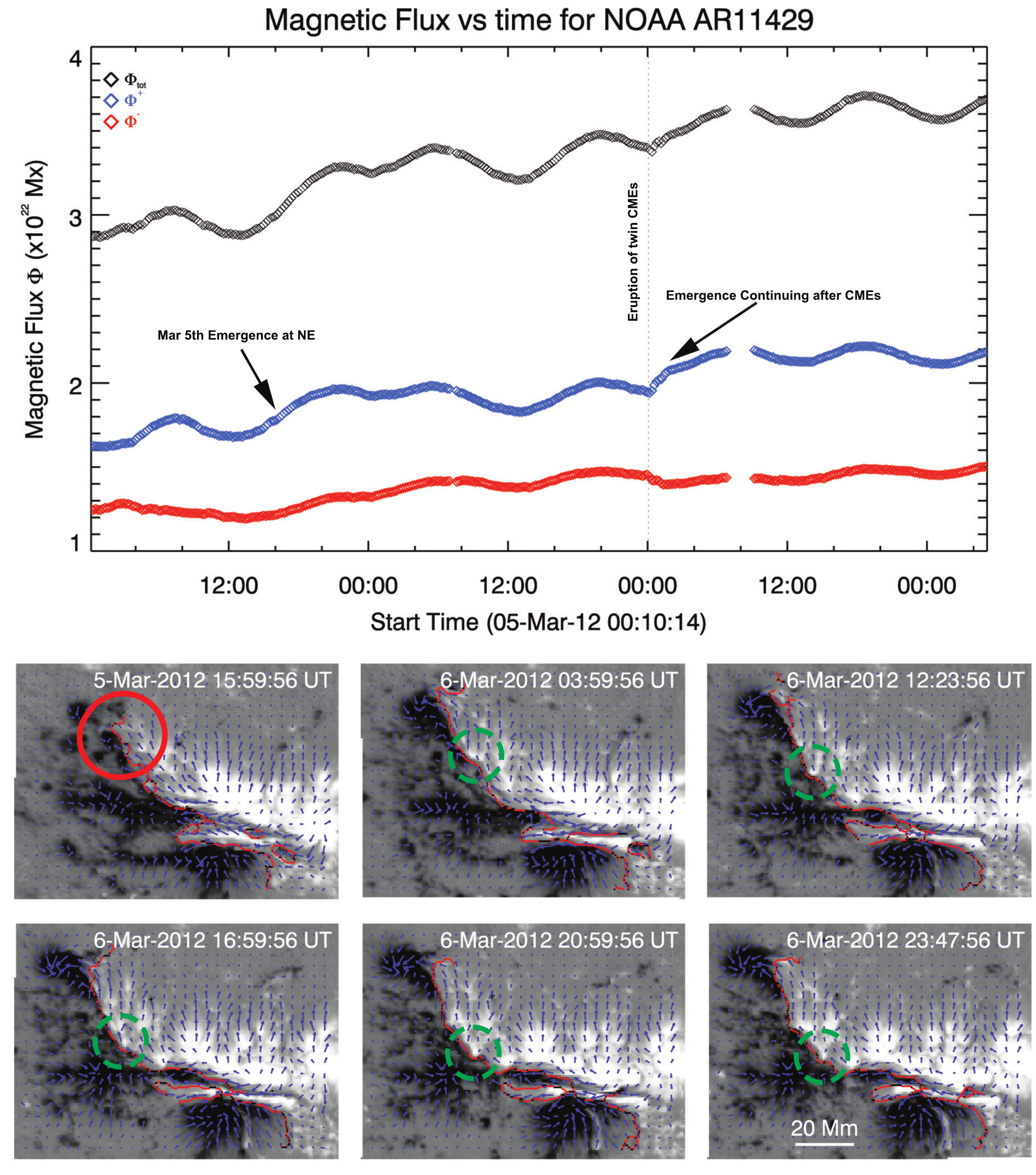} 
\caption{
Top: The evolution of the magnetic flux for AR 11429 between Mar 5 and 8, 2012 (Black diamonds: total unsigned, blue: positive, red: negative). The 1-day period oscillatory behavior is due to intrinsic instrumental problems of the HMI magnetograph becoming obvious in strong-field regions. Nevertheless, the overall trend is increasing in accordance with a continuously developing AR 11429. There are two distinctive flux emergence events showing their imprints in the flux profile - one on Mar 5, 2012 and a second one during/after the eruption of the twin CMEs (shown with arrows).  
Bottom panels: The time evolution of the photospheric vector field during March 5 and 6, 2012, i.e. before the eruptions of March 7, 2012. The blue vectors illustrate the horizontal component of the photospheric magnetic field. With red color we delineate the PIL (obtained by calculating the gradient of the $B_z$ component). Note that while the overall spatial distribution remains the same, significant changes occur along the PIL by means of shearing motions of individual flux elements (green dashed circles; most notably in the NE side of the PIL - white ``blobs'', following an emergence event on March 5, 2012 close to the NE negative polarity; location is shown with a red circle). Shearing occurs at the location of the SW negative sunspot with the elongated positive polarity on top of it. A movie (``movie1.mpg'') is available in the online version of the journal.
}\label{FIG_EVOL}
\end{figure*}

\begin{figure*}
\epsscale{2.0}
\plotone{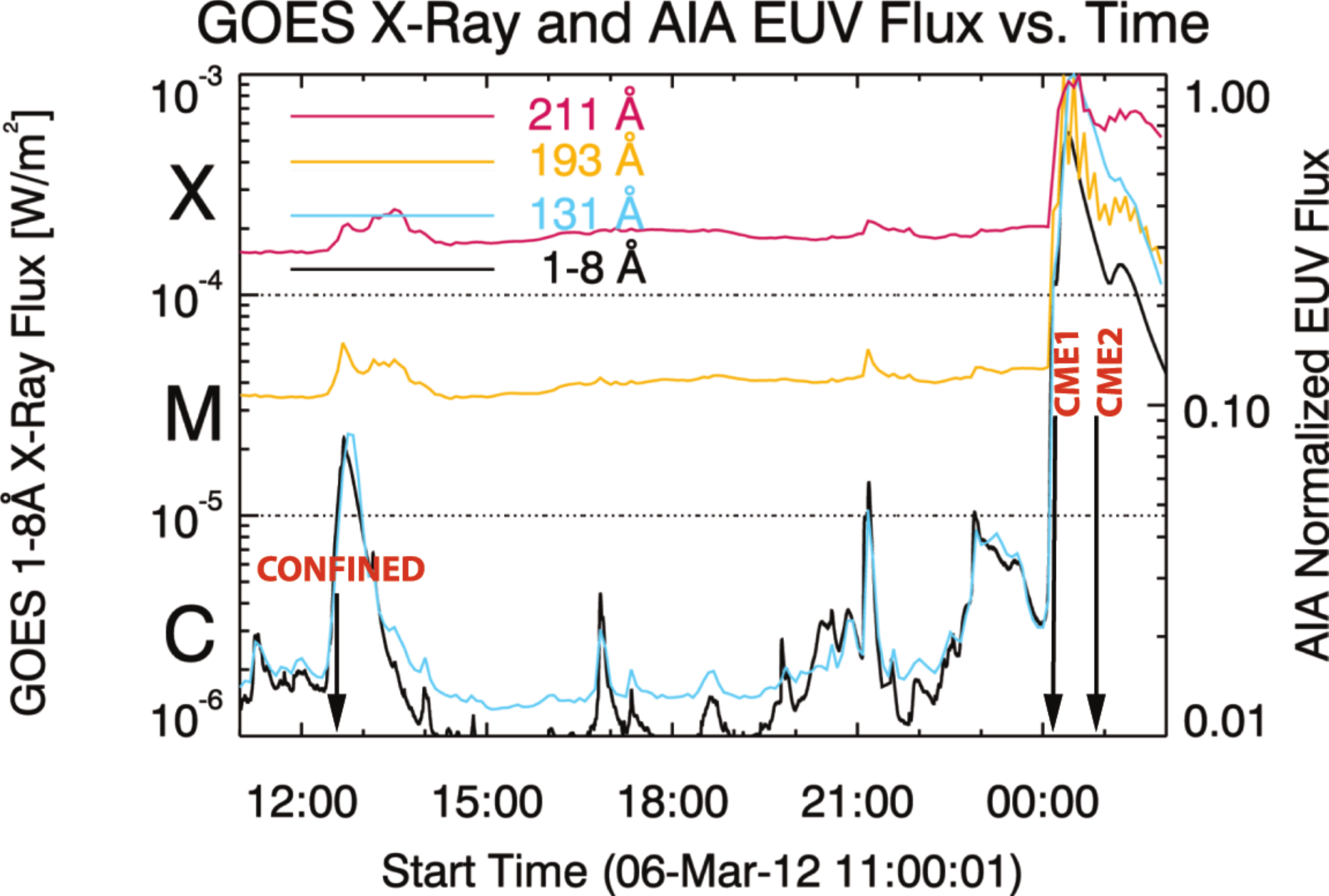} 
\caption{
Soft X-ray flux versus time observed by the GOES satellite (black curve). The EUV normalized flux from  AR 11429 for a few AIA channels is overplotted. The size of the box for the EUV flux calculation is 600$\arcsec \times$500$\arcsec$ (corresponds to the box in Fig~\ref{FIG_AIA}). Note the similarity in the response of the 131\,\AA\ (cyan curve) with the 1-8\,\AA\ from GOES which shows that  AR 11429 is the primary contributor of X-ray flux during the observations.
}\label{FIG_GOES}
\end{figure*}

\begin{figure*}
\epsscale{2.0}
\plotone{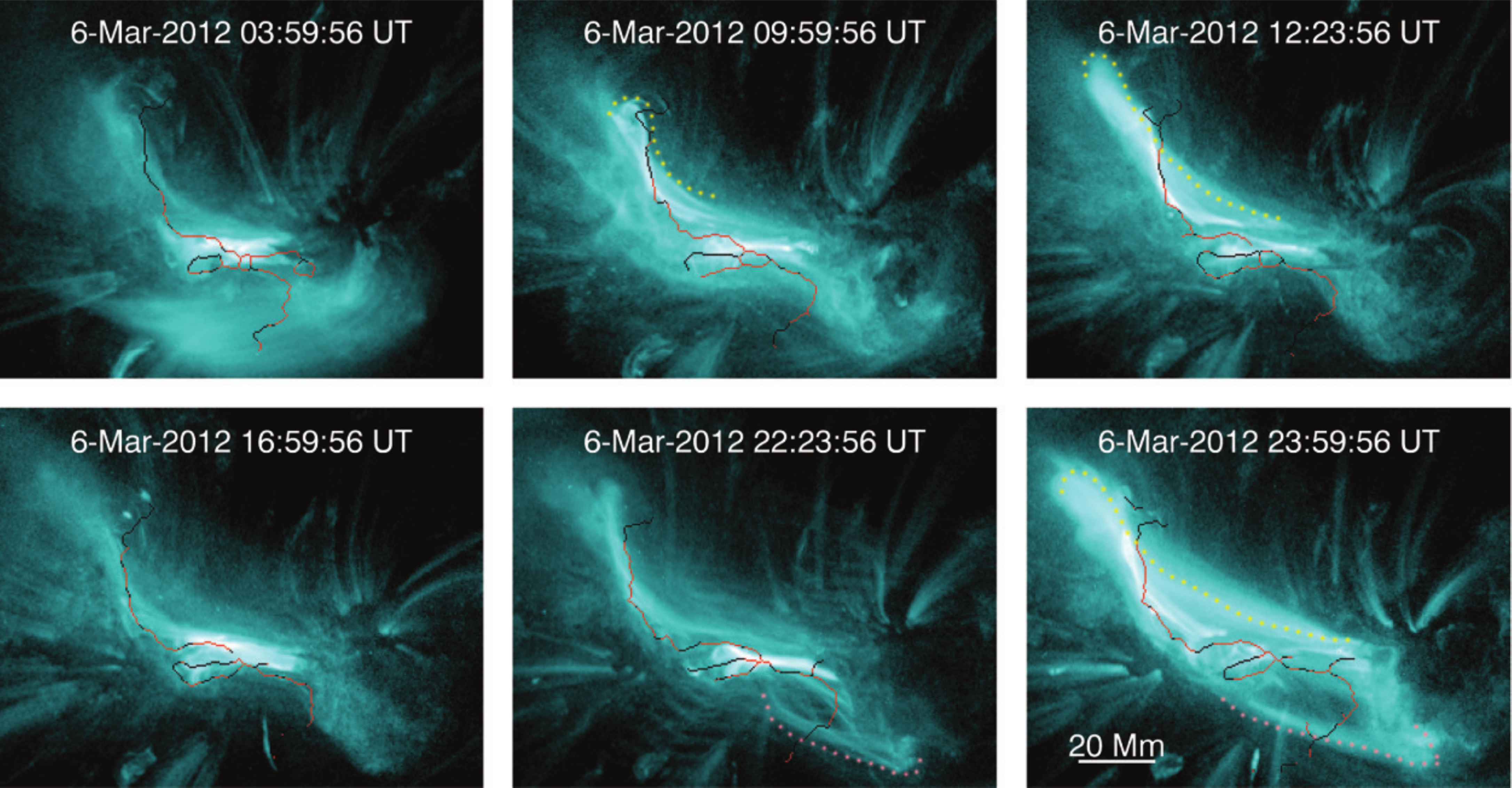} 
\caption{
Time evolution of 131\,\AA\ images from SDO/AIA. The thin super-imposed curve shows the location of the PIL at the photosphere (red shows location where the shear angle between the observed photospheric horizontal field and the potential field as a reference, is greater than 60$\degr$). The PIL was calculated at the CEA Bz frames of  fig. \ref{FIG_EVOL} and it has been transformed into helioprojective to be used with the 131\,\AA\ images. The majority of the transient activity resides at the PIL, and due to the position angle ($\sim$ E30$\degr$N18$\degr$), the transient bright structures are close to the photospheric surface. The MFRs are delineated with a dotted curve; yellow for the NE MFR and pink for the SW. The NE and SW MFR are offset to the PIL, which signifies potentially higher lying structures.
}\label{FIG_AIA_EVOL}
\end{figure*}

\begin{figure*}
\epsscale{2.3}
\plotone{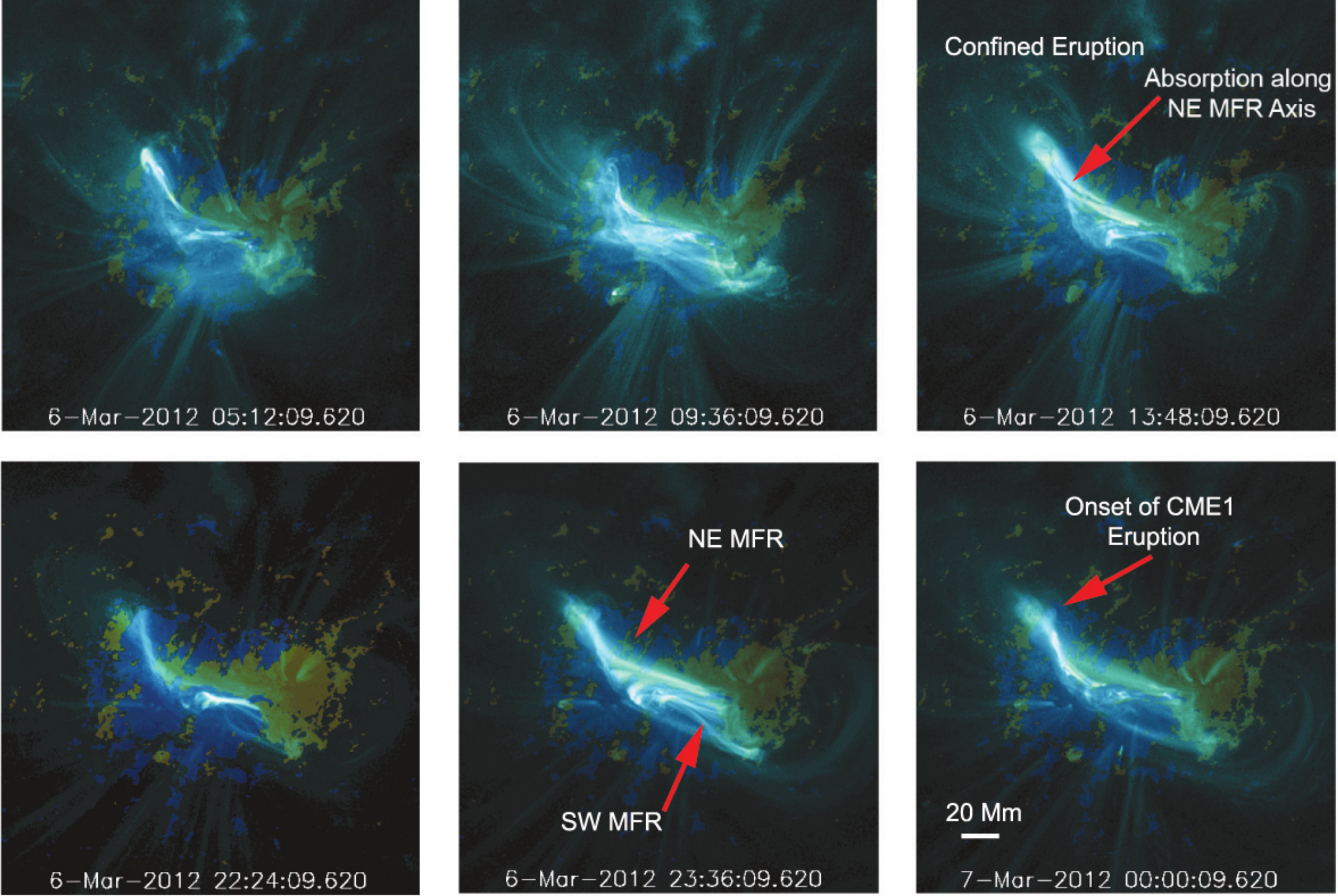} 
\caption{
The evolution of the AR in AIA/131\AA\ with HMI/BLOS composite (gold/positive, navy blue/negative). The locations of the NE MFR and SW MFR are marked by arrows in the middle top and bottom panels. During the cooling phase of the failed eruption ($\sim$ 13:48 UT), an absorption feature appears along the presumed NE MFR axis. The absorption feature is offset from the PIL, which may suggest that it is a high lying formation (due to the geometrical projection from our viewing angle). This absorption feature is also seen in 195\,\AA\ from STEREO B's viewpoint and it is indeed high-lying (see Fig~\ref{FIG_EUVI_ABSORPT}). The associated movie to this figure (``movie2.mpg'') can be found in the online version of the paper. 
}\label{FIG_COMPOSITE_MULTI}
\end{figure*}

\begin{figure*}
\epsscale{2.2}
\plotone{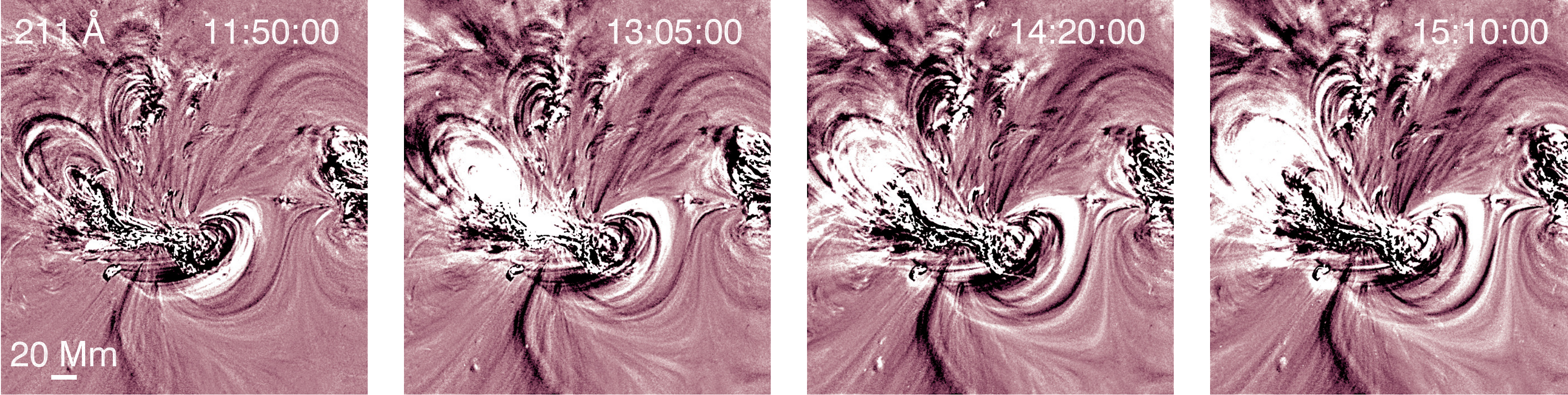} 
\caption{
211\,\AA\ base difference images around the time of the M2.1 flare of March 6, 2012. The absence of spreading dimmings areas in the vicinity of the AR suggest that this was an confined flare.
}\label{FIG_BDIFF}
\end{figure*}

\begin{figure*}
\epsscale{1.9}
\plotone{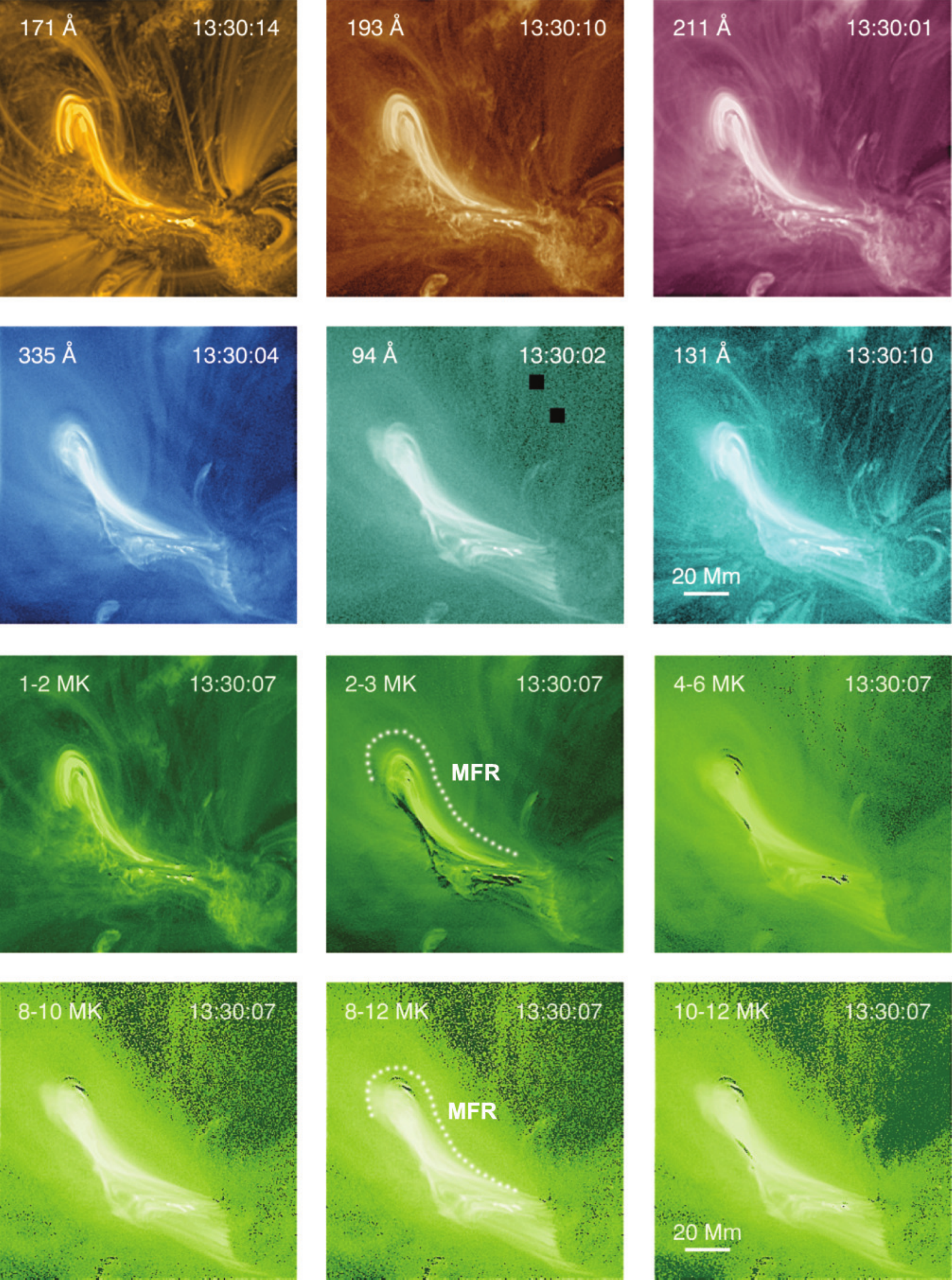} 
\caption{
Top six panels show the MFR candidate structure in the six coronal passbands (131\,\AA, 94\,\AA, 335\,\AA, 211\,\AA, 193\,\AA\ and 171\,\AA) observed one hour after the confined M2.1 flare of March 6, 2012. In the bottom six panels (colored in green) we present DEM images in the displayed temperature ranges. The time of each temperature frame corresponds to the mean time of the coronal passbands (maximum time difference between passband images is 10 sec). The MFR candidate is seen above 2\,MK (dotted line).
}\label{FIG_DEM}
\end{figure*}

\begin{figure*}
\epsscale{2.2}
\plotone{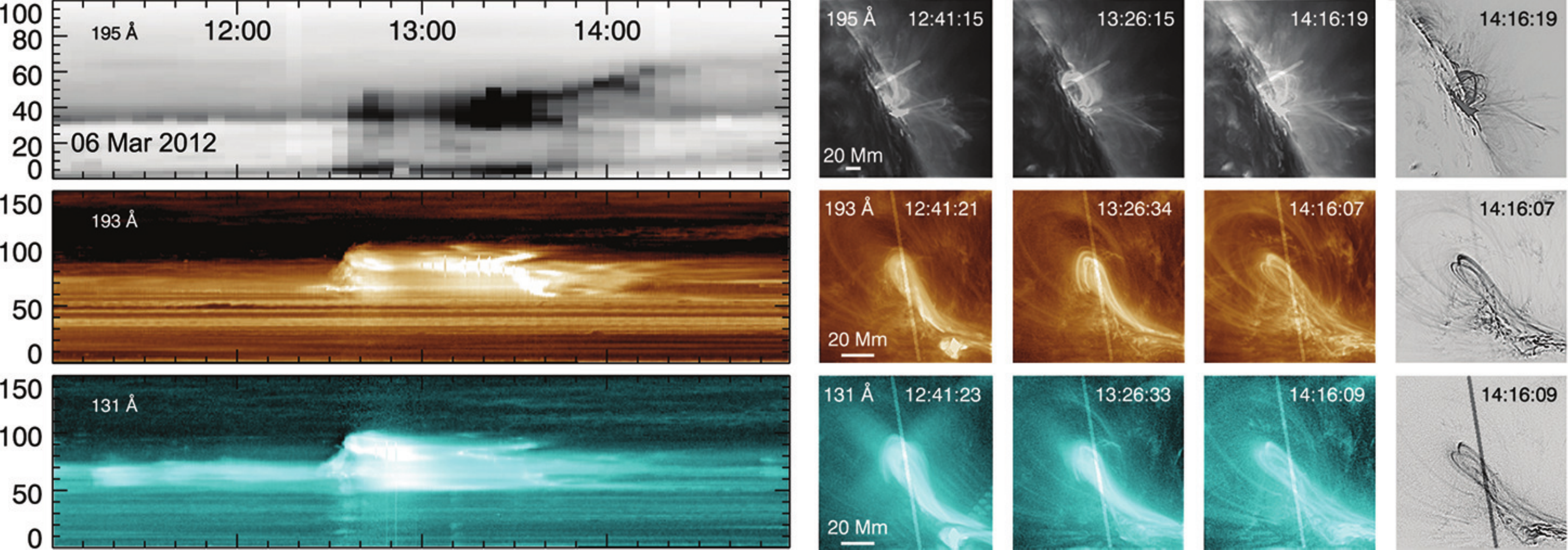} 
\caption{
Left: Slit time plot for EUVI B 195\,\AA\ (top), AIA 193\,\AA\ (middle) and 131\,\AA\ (bottom). The EUVI B slit plot's time cadence is 300 sec and the AIA's is 12 sec. The vertical axis is in arcseconds along the slit. Right: Selected frames showing the location of the slits for EUVI B and AIA images. The slit was selected to intercept the image along the direction of fastest expanding motions. The frames in the last column are laplacian-filtered to enhance the details. A movie is available in the online version of the paper (``movie3.mpg''). 
}\label{FIG_EUVIB}
\end{figure*}

\begin{figure*}
\epsscale{2.0}
\plotone{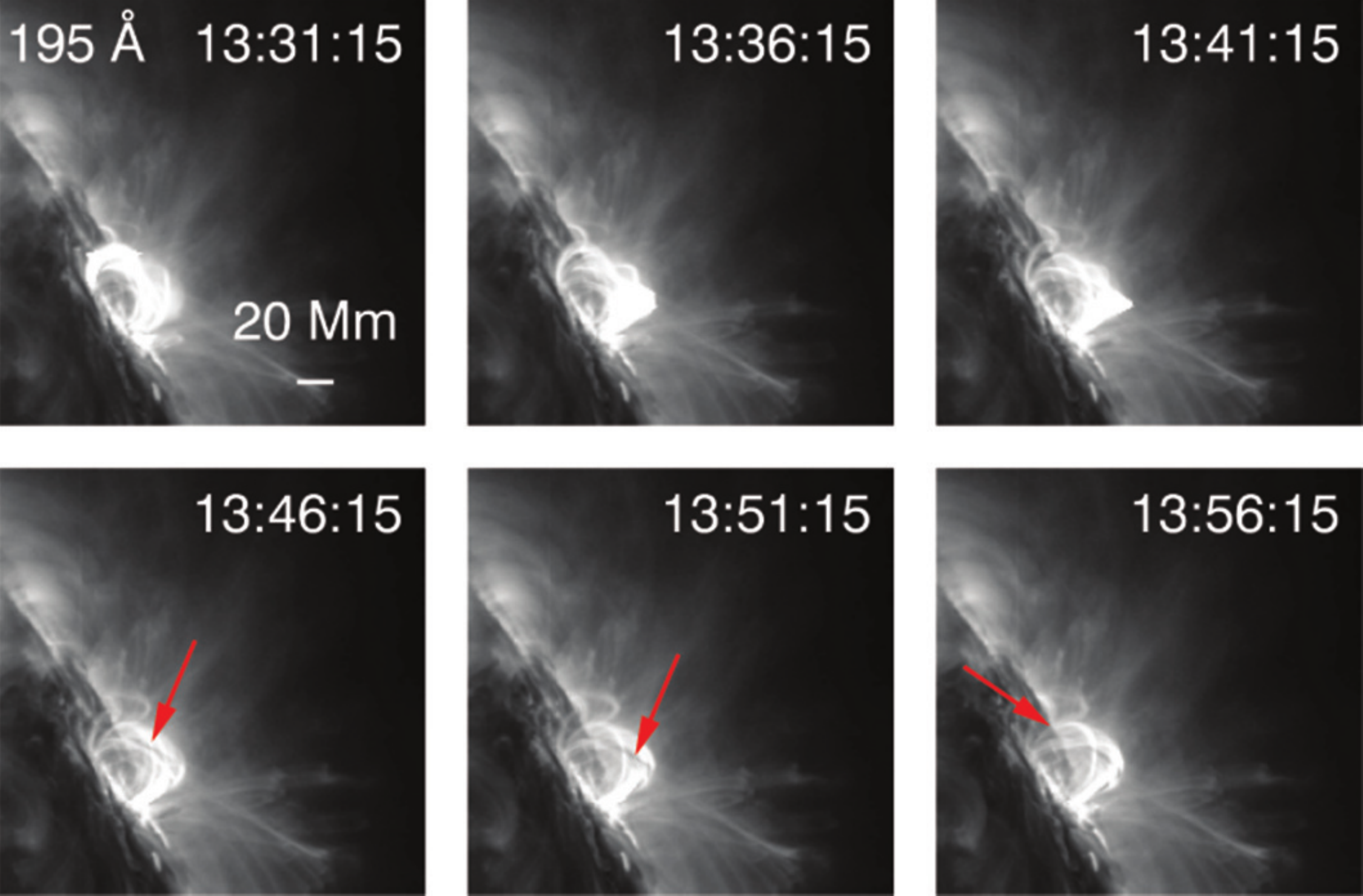} 
\caption{
The final stages of the failed eruption of March 6, 2012 12:30 UT as seen by STEREO B/EUVI at 195\,\AA. Note the absorption feature along the presumed axis of the NE MFR shown with red arrows. This is evidence for a magnetic structure able to hold cool material as expected for the axis of the MFR. The height of the absorption feature is $\sim$30\,Mm. The formation of the absorption feature observed in 131\,\AA\ can be seen in the online movie (``movie3.mpg'').
}\label{FIG_EUVI_ABSORPT}
\end{figure*}

\begin{figure*}
\epsscale{2.2}
\plotone{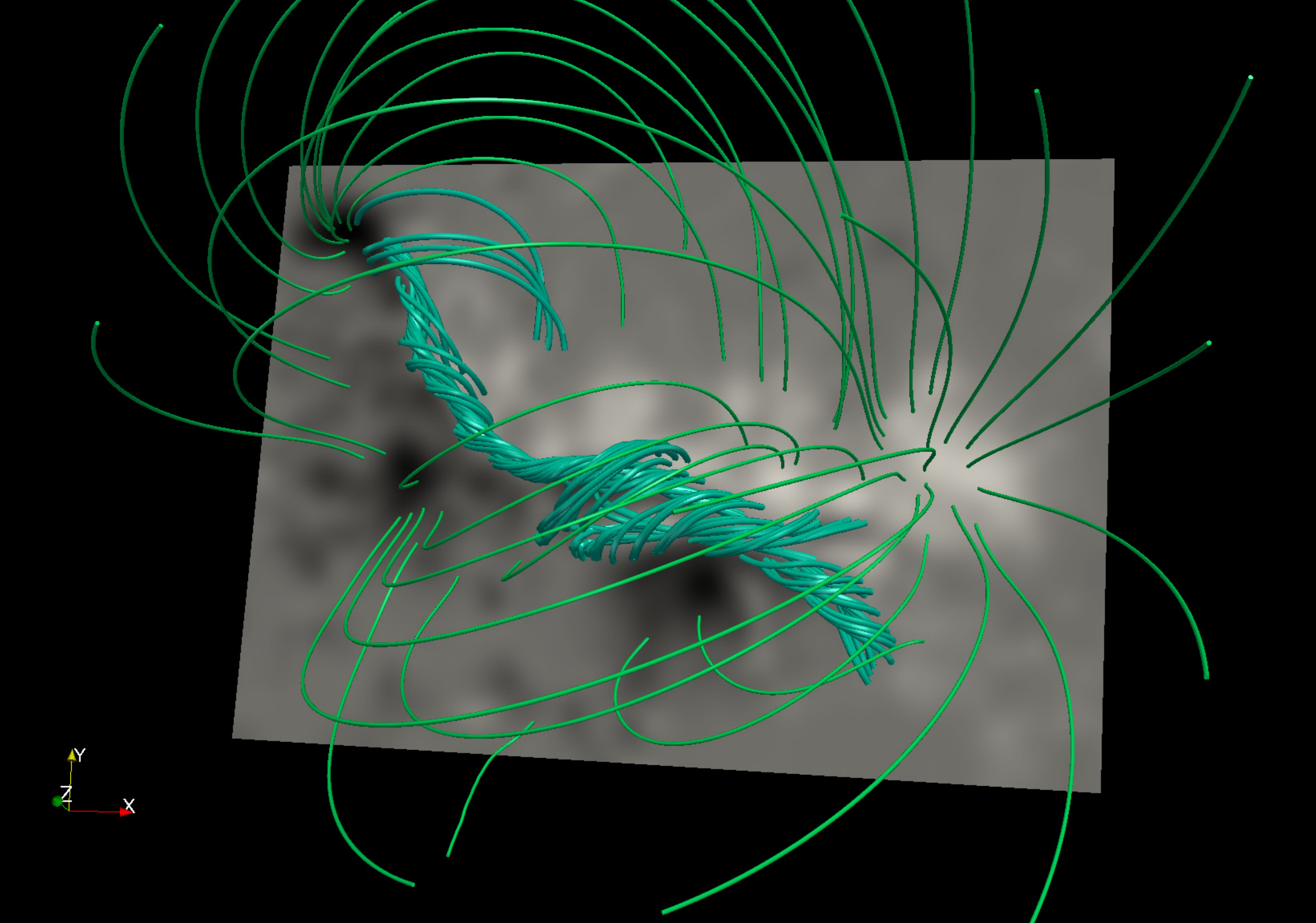} 
\caption{
NLFFF extrapolation of March 6, 2012 23:48 UT. The greyscale map is the $B_z$ component of the photospheric boundary at that time. The colored tubes represent the extrapolated magnetic field in the domain. The teal-colored tubes correspond to the field lines along the PIL. The magnetic field lines are highly sheared which indicates the existence of strong electric currents in the vicinity of the PIL. These field lines are rooted in randomly sampled points within areas of $|\alpha|=5 \times 10^{-9} cm^{-1}$ at the surface. A movie is available in the online version of the paper (``movie4.avi'').
Note the existence of two main ``chains'' of non-potential field lines - a short one, above the negative sunspot in the South and a longer one following the NE PIL. These correspond to the locations of the brightenings seen in 131\,\AA\ and also to the locations of initiation of the two CMEs. The green tubes represent the overlying (nearly potential) field lines sampled randomly in the FOV.
}\label{FIG_EXTRAPOL}
\end{figure*}

\begin{figure*}
\epsscale{2.2}
\plotone{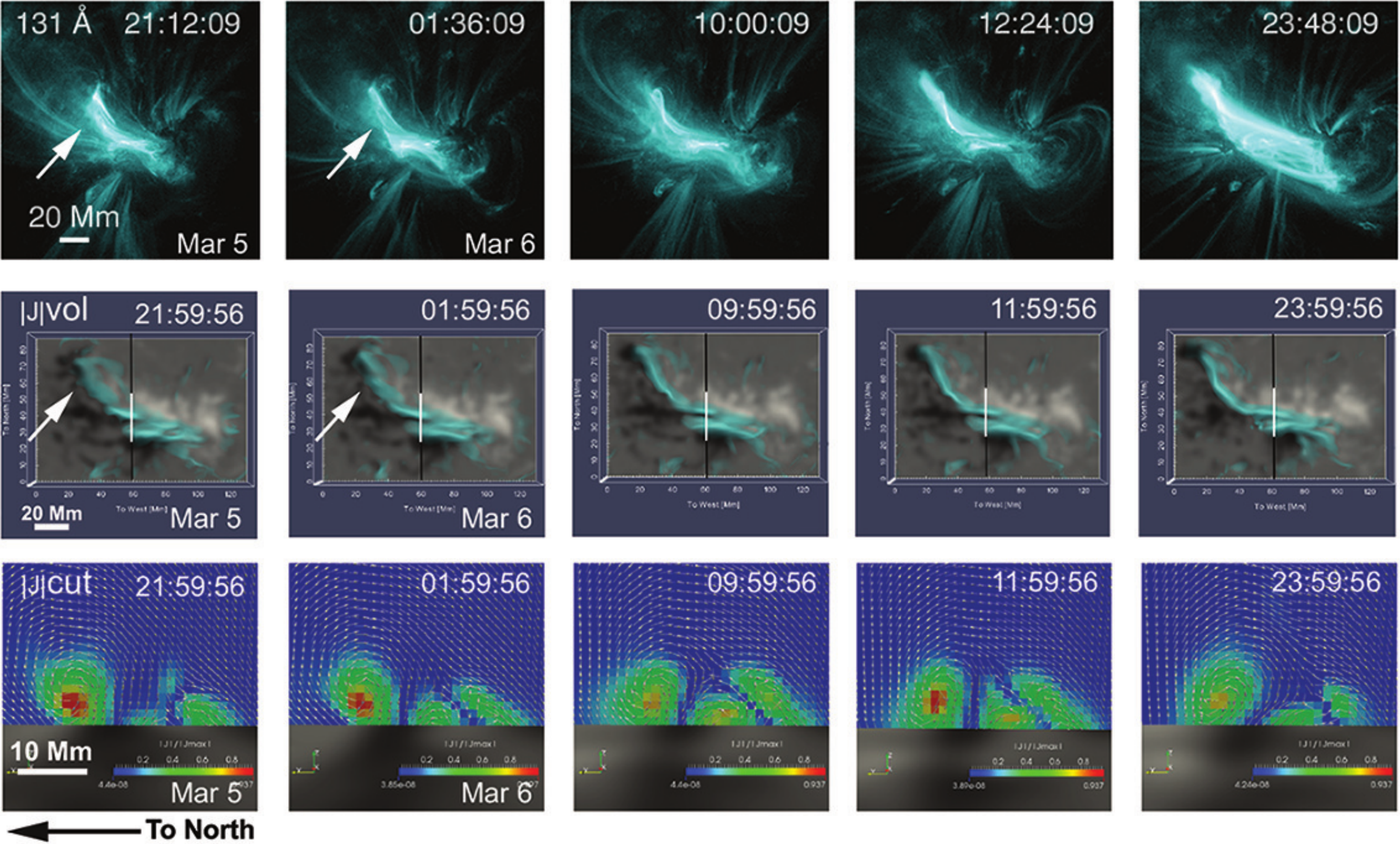} 
\caption{
Top row: 131\,\AA\ images showing the evolution in coronal plasma emission of AR11429 at 10\,MK. Middle row: Volumetric visualization of the magnitude of strong (normalized to peak current density) electrical currents calculated via Amp\`{e}re's law in the NLFFF volumes. The FOV roughly corresponds to that of the top row; the volume structure is seen from above. The volumetric currents are represented in an ``optically-thin'' fashion and a teal color table has been used for the visualization. These strong currents reside close to the surface as also seen in the vertical plane cuts (see bottom row). Bottom row: Vertical cuts of the 3-D volume $|\mathbf{J}|$ at the position indicated in the middle row (black line). Color contours show strong $|\mathbf{J}|$ with red and weak with blue. The magnetic field cut by the vertical plane is overplotted (note the strong curl in regions of high $|\mathbf{J}|$). The greyscale maps correspond to the photospheric boundary. Transient brightenings in the 10\,MK (131\,\AA) channel seem to be in good agreement with  $|\mathbf{J}|$-volume channels inferred from the NLFFF. The NE PIL appears fragmented at earlier times (white arrows) and progressively straightens, lengthens and becomes an integral structure as signified by the transient brightenings. Several hours prior to eruption, two main hot $|\mathbf{J}|$-channels are visible, in agreement with 131\,\AA\ .
}
\label{FIG_MULTI_J}
\end{figure*}

\begin{figure*}
\epsscale{1.5}
\plotone{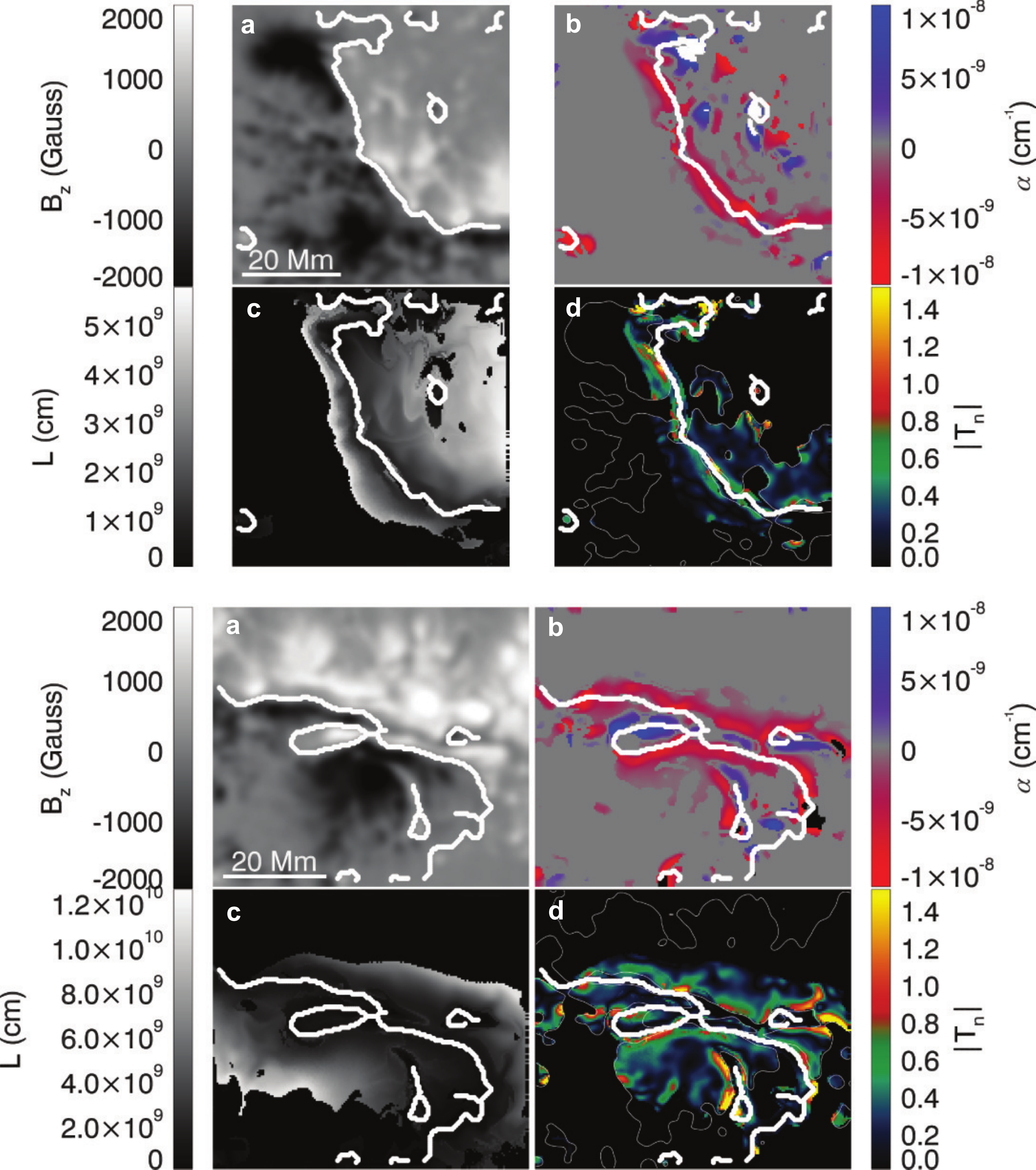} 
\caption{Helicity-related parameters from the extrapolation of March 6, 2012 23:46 UT (i.e., the same shown in Fig~\ref{FIG_EXTRAPOL}). Top four panels: NE of fig~\ref{FIG_HMI}): (a) Bz map, (b) force-free $\alpha$ map, (c) map of field line lengths, L, and, (d) the unsigned twist number $|T_n|$. PIL is shown with a white curve. Bottom four panels: Same parameters but for the SW box of Fig~\ref{FIG_HMI}. Only a small fraction of pixels in the vicinity of the PIL exhibits significant twist ($|T_n|>0.5$). }
\label{FIG_TWIST_MAP}
\end{figure*}

\begin{figure*}
\epsscale{1.4}
\plotone{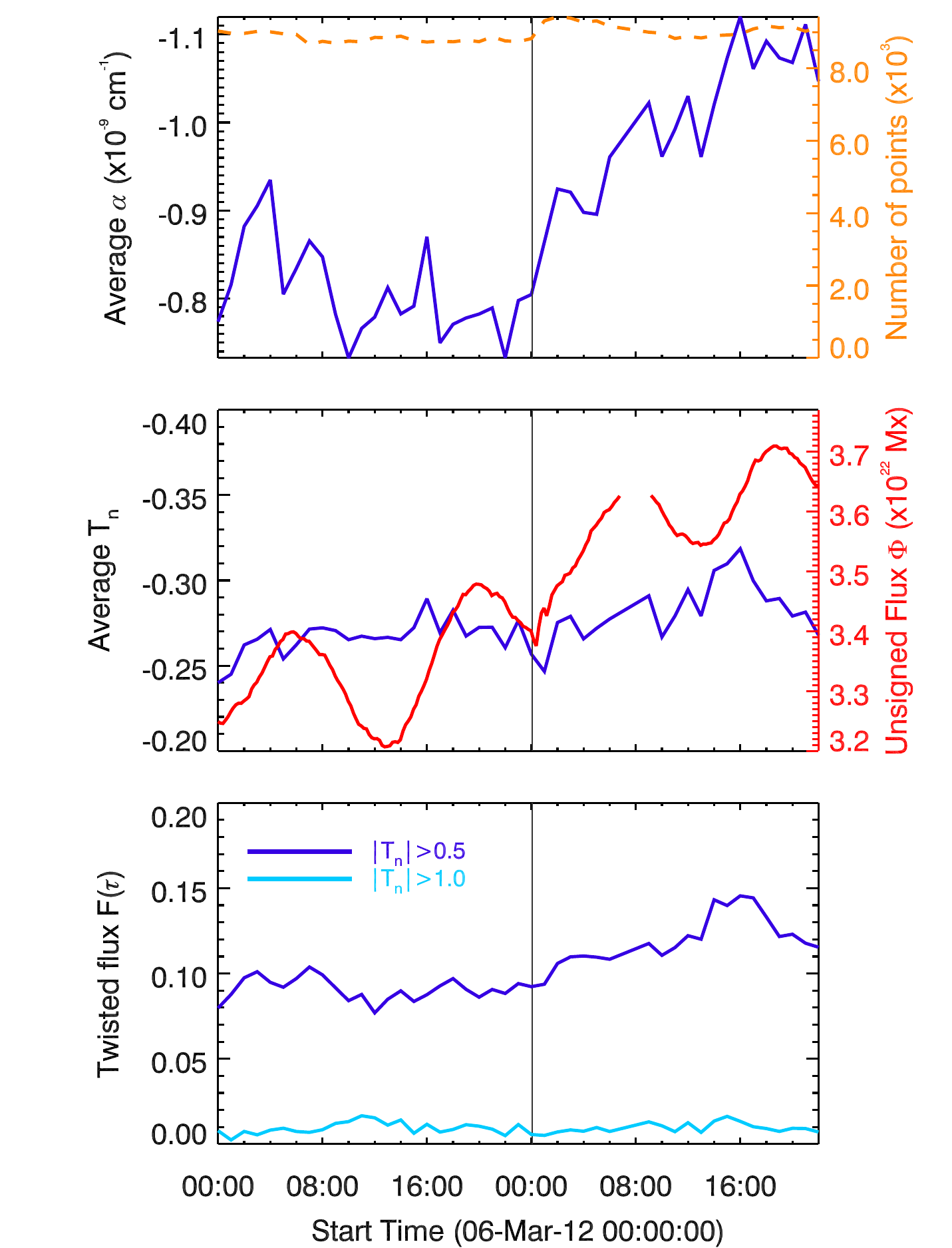} 
\caption{
The 2-day evolution of the average force-free $\alpha$ (top), average twist number (middle) and the fraction of flux with twist higher than a twist threshold (bottom). Note that the y-axes for top and middle panels are reversed. Only a fraction of the vicinity of the PIL exhibits significant twist. The twisted flux increases after the two CMEs (vertical line), which suggests continuing flux emergence past the time of the CME eruptions. The unsigned magnetic flux is shown in the middle panel (red). Note the fast increase of magnetic flux right after the eruptions. Recall that our results are at 1-hour cadence, which compares to the temporal lag between the onsets of the two CMEs.
}
\label{FIG_TWIST}
\end{figure*}

\begin{figure*}
\epsscale{2.1}
\plotone{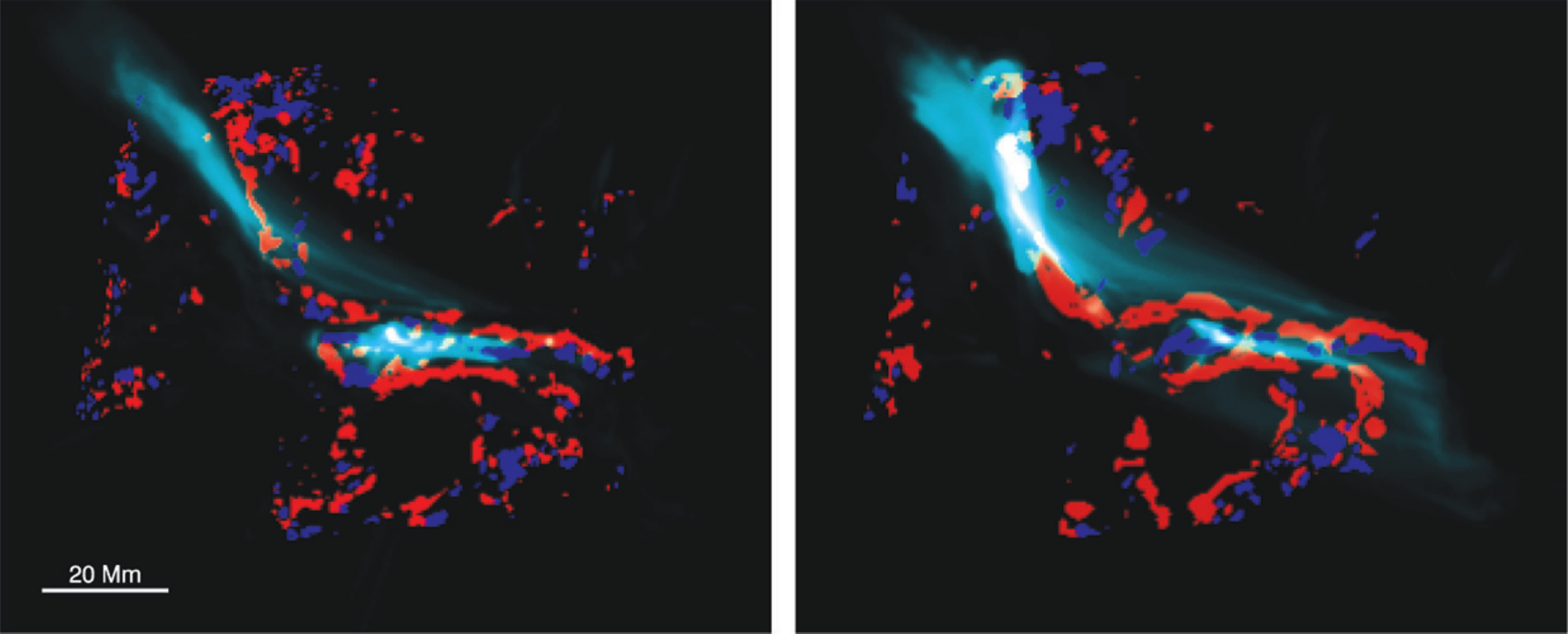} 
\caption{
Composites of 131\AA\ observations and force-free $\alpha$-maps calculated at the photosphere. Left panel shows the map at March 6, 2012, 10:59:56 UT and on the right at March 6, 2012, 23:47:56 UT. Blue/red color corresponds to positive/negative $\alpha$. The teal-colored overlaid map corresponds to the $\sim$2-hour accumulated brightenings (up to 2-hours before the aforementioned times) in the 131\,\AA\ AIA passband by calculating the standard deviation of each pixel in the 131\,\AA\ image time series. Note that the $\alpha$-maps are converted back to helioprojective coordinates to match the AIA standard deviation maps. 
}\label{FIG_ALPHA}
\end{figure*}

\begin{figure*}
\epsscale{2.1}
\plotone{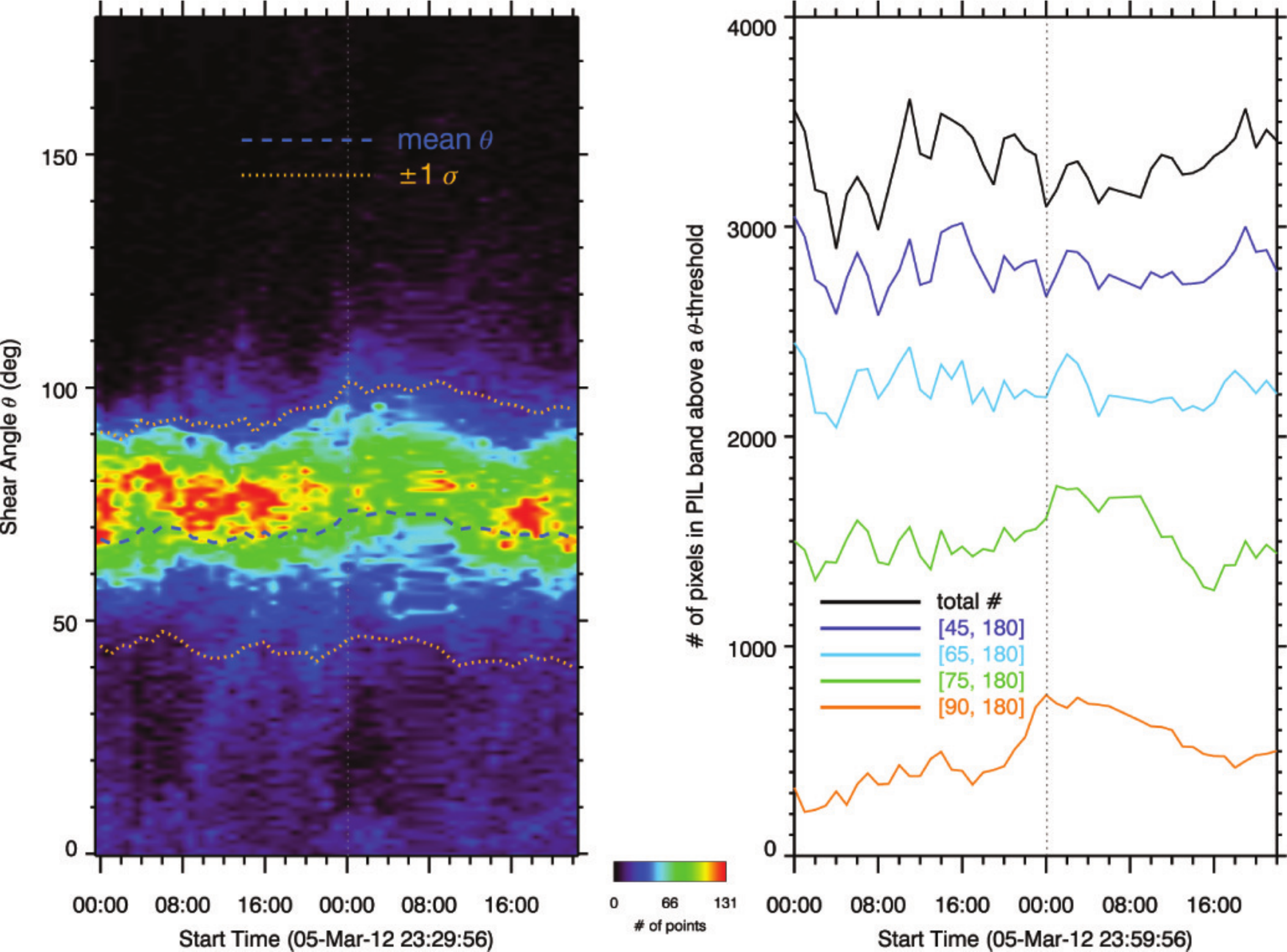} 
\caption{
Left: The distribution of the non-potential shear angle $\theta$, i.e. the angle between the horizontal component of the observed photospheric field, $\mathbf{B}_{obs}$, and the field from the potential extrapolation,  $\mathbf{B}_{pot}$, at an hourly step, for 2 days (each column of the left panel of this figure corresponds to the theta angle histogram  at the corresponding time). The mean shear builds up $\sim$ 4 hours prior to the X5.4 flare of 00:02 UT March 7, 2012 and the 2$\sigma$-range of shear shifts with the mean shear. The fact that the shear is relatively strong before and after the CMEs, may suggest that part of it is involved in the MFR formation. Right: Total number of pixels in the PIL band above a $\theta$-threshold. For $\theta\geq$75$\degr$ the increase of the number of highly sheared PIL pixels seems to occur $\sim$ 8 hours prior to the eruption. 
}
\label{FIG_SHEAR}
\end{figure*}

\begin{figure*}
\epsscale{1.1}
\plotone{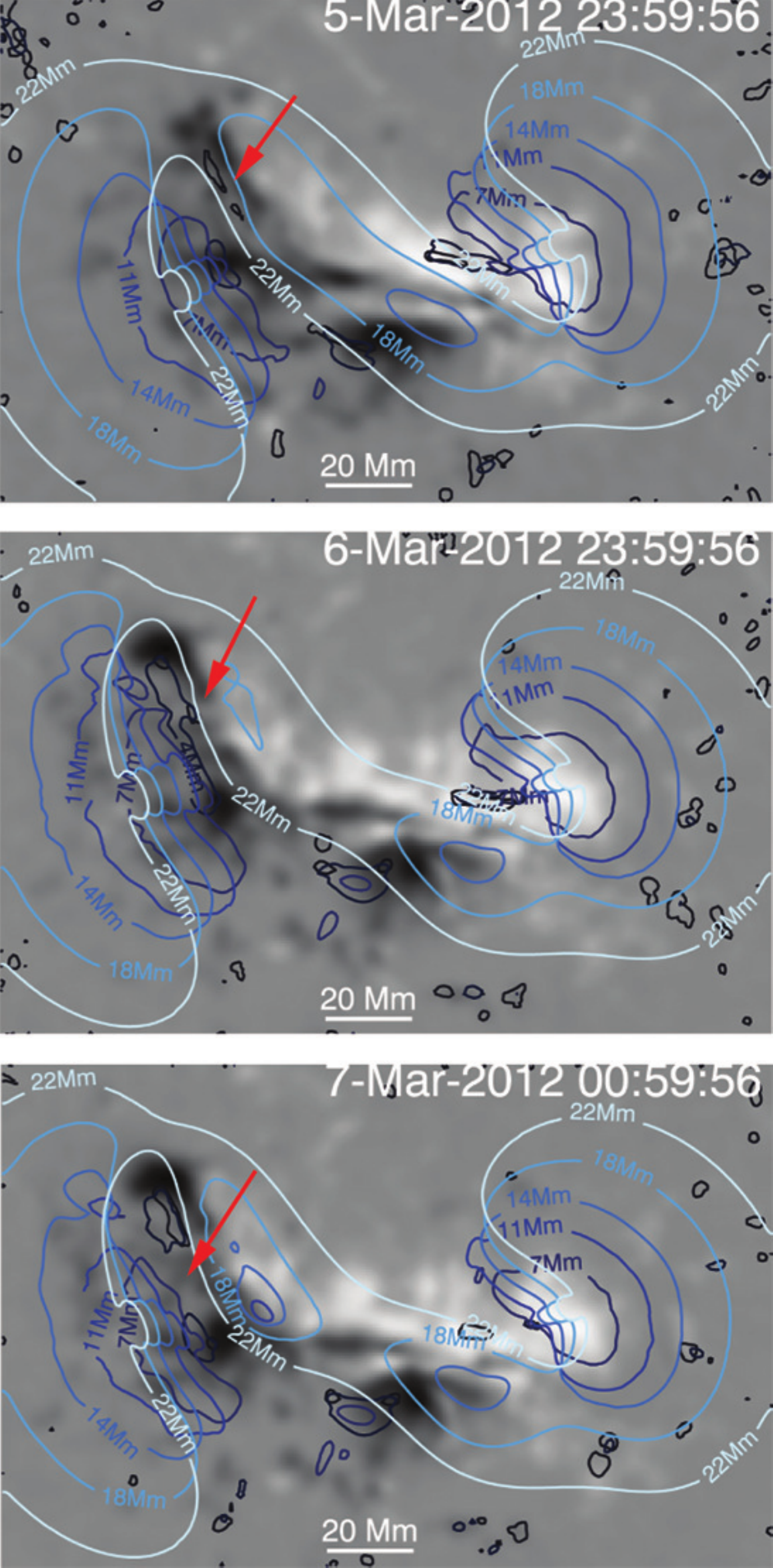} 
\caption{
Decay index with height contour plot overlaid on photospheric CEA Bz maps (color contours correspond to decay index n=1.5 at each height). The decay index profile is calculated for each z-column of the horizontal magnetic field from the NLFFF cubes. Top panel is at the beginning of the observations, middle is a few minutes prior to CME1 and bottom right before CME2. Areas within color-contours contain ``super-critical'' decay indices at each height, $n\geq1.5$. Note the activity in the low-lying (darker colors) contours of the NE PIL. In just one day (top frame to middle) an elongated and continuous stacking of ``super-critical'' decay-index areas is being formed, giving birth to a ``super-critical'' decay-index ``tunnel'' (see succession of closed contours at the location pointed by the red arrows). The succession of these super-critical decay-index ``holes'' signifies a least-resistance path, or the most probable path for the CME1 initial trajectory (i.e. towards the East). A movie is available in the online version of the paper (``movie5.mpg'').
}
\label{FIG_DECAY_INDEX}
\end{figure*}

The detailed NLFFF extrapolations allowed to estimate the poloidal flux $\Phi_p$ of the MFRs before the CME eruptions. For the NE MFR (associated with CME1)  $\Phi_p \sim$5.82$\times\,10^{20}$\,Mx and for the SW MFR (CME2) $\Phi_p \sim$3$\times\,10^{21}$\,Mx. The SW MFR, although shorter, showed higher spatial frequencies and thus more crossings in the calculation plane per unit length. It was also rooted at stronger magnetic fields. The height for the calculation was determined to be up to 15\,Mm since any further increase in height did not change the values considerably. This was consistent with our finding of low-lying MFR structures (Section~\ref{stream_line_section}). Unfortunately, the MC fittings for CME2 were highly uncertain so we were unable to derive an 1 AU poloidal flux to compare with our origin estimates. Comparing our fluxes with statistics of MCs observed at 1\,AU we find good agreement (e.g., $\Phi_p \sim$ 5$\times\,10^{20}$ -- 2$\times\,10^{22}$ \,Mx,  (Figure~8, \citealt{Qiu_etal_2007}), $\Phi_p \sim$ 8$\times\,10^{20}$ -- 1.6$\times\,10^{21}$ \,Mx (Table 1, \citealt{Moestl_etal_2009})). In particular, our value for the poloidal flux seems to be at the center of the statistical distribution of poloidal fluxes from Fig. 2 of \citet{Lynch_etal_2005}. We surmise that a significant amount of poloidal magnetic flux possibly existed in the MFR before its eruption. Overall, the combination of the solar observations and detailed NLFFF modeling enabled us to detect the MFR structures at least 12 hours before the eruptions. We summarize our major findings in the following list:
  
\begin{enumerate}
\item The AR contains two major stressed magnetic flux rope systems (MFRs), one in the NE and the other in the SW. We propose that the NE MFR (giving rise to CME1) formed during major confined flares, while the SW MFR (giving rise to CME2) formed during confined sub-flaring episodes.

\item We observe \textit{strong shearing} motions. Oppositely oriented magnetic elements move  close to each other and give rise to sharp, compact polarity inversion lines (PILs); one PIL extends along the NE and another, shorter, PIL extends along the SW of the active region (Fig~\ref{FIG_HMI} \& online movies ``movie1.mpg'', ``movie2.mpg''). These shearing motions last throughout our observing period and eventually lead to a non-potential field topology. 

\item The strong shear leads to MFR formation via \textit{reconnection events}  during confined flaring episodes, ranging from proper flares to sub-flares. The flare heating "illuminates" the newly created MFR field lines in hot AIA channels (e.g., standard-deviation maps of Fig~\ref{FIG_ALPHA}).  

\item We find that a similar number of pixels in the vicinity of the PIL has both high twist ($T_n \geq 0.5$) and high shear ($\theta \geq 75 \degr$). This suggests weakly twisted non-potential structures along the PIL.

\item Further support that the NE MFR forms during the confined flare on March 6, is offered by the appearance of cool filamentary plasma along the presumed MFR axis during the cooling phase ($\sim$13:49:21). This indicates the existence of a coherent structure along the PIL, able to support cool plasma (in agreement with MFR observations of \citet{Patsourakos_etal_2013} that show cooling at the MFR center; see also ``movie3.mpg'').

\item The pre-eruptive MFRs are low-lying and \textit{weakly-twisted}. This is a challenge for their detection due to background emission in the lower corona. However we see them during flaring events when they do clearly stand out from the background. 

\item  The low twist, seen in the EUV observations and produced in the NLFFF extrapolations, favors the ideal-MHD torus instability as the driver of the eruption.

\item The two CMEs originate from two PILs within the same active region. There is evidence for pre-existing MFRs over both PILs. The two PILs are two different flux systems and the formation of the two MFRs reflects that duality. The NE MFR is formed via a confined flare $\sim$ 12 hours prior to its CME (e.g. similar structure in right column of Fig~\ref{FIG_AIA_EVOL} 12 hours apart; yellow dotted line). The SW MFR forms via small-scale flaring over many hours. The two CMEs continue to express their different nature even during the eruption where they follow paths determined by the gradient of the magnetic field above each MFR. 
\end{enumerate}

In conclusion, we were able to use detailed observations from SDO and STEREO and sophisticated NLFFF extrapolations to decipher the origins of the double CME event that occurred on March 7, 2012. They erupted from two different PIL systems, within the same active region. By tracing the coronal activity along those PILs backwards in time, we found strong evidence for the formation of magnetic flux ropes during confined flaring events of various magnitudes
ranging from a proper flare (the M-class flare on March 6 at around 12:35~UT) down to sub-flare events.

Hot ($\sim$ 10 MK) elongated EUV structures appeared at those times.
Their behavior was consistent with an MFR seen face-on considering past observations of MFR detections on the limb. NLFF extrapolations based on HMI observations over the full two-day period supported the EUV observations.  They indicated the gradual formation of low-lying, weakly twisted structures at the same locations where the hot EUV signatures were detected. Furthermore, the evolution of the large scale field structure around the AR exhibited two areas of sharp gradients in the  magnetic field magnitude as a function of height. Such locations are considered the 'pathways' through CMEs tend to erupt in the low corona since they provide the path of least resistance for the erupting system. True to this modeling result, both CMEs on March 7, 2012 followed their corresponding 'pathway', which explains their widely different propagation directions. In short, the two CMEs on March 7, 2012 were the eruption of magnetic flux ropes formed 12 hours earlier during a confined flare episode (for CME1) or over several hours via sub-flaring activity (for CME2). In other words, the two CMEs should be considered as separate events because they originate from different flux systems even though the two systems belong to a single AR. It raises the question on whether such complex $\delta$-spot regions should be considered as single ARs or not.

Finally, we note that the sheer complexity of the photospheric and coronal configuration of the studied AR presented a formidable challenge for the data analysis and NLFFF modeling. However, such complex, but rare, ARs, are the most dangerous in terms of space weather conditions and their study is therefore vitally important in understanding the solar conditions that lead to extreme eruptive activity.


\acknowledgments
The authors are grateful to the anonymous referee for the careful review and constructive comments. The authors wish to thank Drs. M. Linton, B. Kliem, and M.K. Georgoulis for valuable discussions. The authors also wish to thank Dr. T. Wiegelmann for providing the NLFFF extrapolation code. S.P. acknowledges support from ​an FP7 Marie Curie Grant (FP7-PEOPLE-2010-RG/268288). S.P. acknowledges support from the European Union (European Social Fund – ESF) and Greek National funds through the Operational Program "Education and Lifelong Learning" of the National Strategic Reference Framework (NSRF) - Research Funding Program: Thales. Investing in knowledge society through the European Social. G. Chintzoglou also thanks Prof. C.E. Alissandrakis, A. Nindos and J. Zhang for inspiring conversation and encouragement.  G.C. was supported by NASA Headquarters under the NASA Earth and Space Science Fellowship Program - Grant NNX12AL73H. A.V was supported by NASA contract S-136361-Y to NRL and internal APL funds.

%


\begin{thebibliography}{64}
\expandafter\ifx\csname natexlab\endcsname\relax\def\natexlab#1{#1}\fi

\bibitem[{{Ambastha} {et~al.}(1993){Ambastha}, {Hagyard}, \&
  {West}}]{Ambastha_etal_1993}
{Ambastha}, A., {Hagyard}, M.~J., \& {West}, E.~A. 1993, \solphys, 148, 277

\bibitem[{{Antiochos} {et~al.}(1999){Antiochos}, {DeVore}, \&
  {Klimchuk}}]{Antiochos_etal_1999}
{Antiochos}, S.~K., {DeVore}, C.~R., \& {Klimchuk}, J.~A. 1999, \apj, 510, 485

\bibitem[{{Archontis} {et~al.}(2009){Archontis}, {Hood}, {Savcheva}, {Golub},
  \& {Deluca}}]{Archontis_etal_2009}
{Archontis}, V., {Hood}, A.~W., {Savcheva}, A., {Golub}, L., \& {Deluca}, E.
  2009, \apj, 691, 1276

\bibitem[{{Archontis} \& {T{\"o}r{\"o}k}(2008)}]{Archontis_Torok_2008}
{Archontis}, V., \& {T{\"o}r{\"o}k}, T. 2008, \aap, 492, L35

\bibitem[{{Aschwanden}(2004)}]{Aschwanden_2004}
{Aschwanden}, M.~J. 2004, {Physics of the Solar Corona. An Introduction}
  (Praxis Publishing Ltd)

\bibitem[{{Aulanier} {et~al.}(2010){Aulanier}, {T{\"o}r{\"o}k}, {D{\'e}moulin},
  \& {DeLuca}}]{Aulanier_etal_2010}
{Aulanier}, G., {T{\"o}r{\"o}k}, T., {D{\'e}moulin}, P., \& {DeLuca}, E.~E.
  2010, \apj, 708, 314

\bibitem[{{Berger} \& {Prior}(2006)}]{Berger_Prior_2006}
{Berger}, M.~A., \& {Prior}, C. 2006, Journal of Physics A Mathematical
  General, 39, 8321

\bibitem[{{Bobra} {et~al.}(2008){Bobra}, {van Ballegooijen}, \&
  {DeLuca}}]{Bobra_etal_2008}
{Bobra}, M.~G., {van Ballegooijen}, A.~A., \& {DeLuca}, E.~E. 2008, \apj, 672,
  1209

\bibitem[{{Canfield} {et~al.}(1999){Canfield}, {Hudson}, \&
  {McKenzie}}]{Canfield_etal_1999}
{Canfield}, R.~C., {Hudson}, H.~S., \& {McKenzie}, D.~E. 1999, \grl, 26, 627

\bibitem[{{Canou} {et~al.}(2009){Canou}, {Amari}, {Bommier}, {Schmieder},
  {Aulanier}, \& {Li}}]{Canou_etal_2009}
{Canou}, A., {Amari}, T., {Bommier}, V., {et~al.} 2009, \apjl, 693, L27

\bibitem[{{Cheng} {et~al.}(2013){Cheng}, {Zhang}, {Ding}, {Liu}, \&
  {Poomvises}}]{Cheng_etal_2013}
{Cheng}, X., {Zhang}, J., {Ding}, M.~D., {Liu}, Y., \& {Poomvises}, W. 2013,
  \apj, 763, 43

\bibitem[{{Cheng} {et~al.}(2011){Cheng}, {Zhang}, {Liu}, \&
  {Ding}}]{Cheng_X_etal_P2_2011}
{Cheng}, X., {Zhang}, J., {Liu}, Y., \& {Ding}, M.~D. 2011, \apjl, 732, L25+

\bibitem[{{D{\'e}moulin} {et~al.}(2013){D{\'e}moulin}, {Dasso}, \&
  {Janvier}}]{Demoulin_etal_2013}
{D{\'e}moulin}, P., {Dasso}, S., \& {Janvier}, M. 2013, \aap, 550, A3

\bibitem[{{Fan}(2009)}]{Fan_2009}
{Fan}, Y. 2009, Living Reviews in Solar Physics, 6, 4

\bibitem[{{Forbes} \& {Priest}(1995)}]{Forbes_Priest_1995}
{Forbes}, T.~G., \& {Priest}, E.~R. 1995, \apj, 446, 377

\bibitem[{{Georgoulis} {et~al.}(2012){Georgoulis}, {Titov}, \&
  {Miki{\'c}}}]{Georgoulis_etal_2012}
{Georgoulis}, M.~K., {Titov}, V.~S., \& {Miki{\'c}}, Z. 2012, \apj, 761, 61

\bibitem[{{Green} \& {Kliem}(2009)}]{Green_Kliem_2009}
{Green}, L.~M., \& {Kliem}, B. 2009, \apjl, 700, L83

\bibitem[{{Green} {et~al.}(2011){Green}, {Kliem}, \&
  {Wallace}}]{Green_etal_2011}
{Green}, L.~M., {Kliem}, B., \& {Wallace}, A.~J. 2011, \aap, 526, A2

\bibitem[{{Guo} {et~al.}(2010){Guo}, {Schmieder}, {D{\'e}moulin}, {Wiegelmann},
  {Aulanier}, {T{\"o}r{\"o}k}, \& {Bommier}}]{Guo_etal_2010}
{Guo}, Y., {Schmieder}, B., {D{\'e}moulin}, P., {et~al.} 2010, \apj, 714, 343

\bibitem[{{Hagyard} {et~al.}(1984){Hagyard}, {Teuber}, {West}, \&
  {Smith}}]{Hagyard_etal_1984}
{Hagyard}, M.~J., {Teuber}, D., {West}, E.~A., \& {Smith}, J.~B. 1984,
  \solphys, 91, 115

\bibitem[{{Hoeksema} {et~al.}(2014){Hoeksema}, {Liu}, {Hayashi}, {Sun},
  {Schou}, {Couvidat}, {Norton}, {Bobra}, {Centeno}, {Leka}, {Barnes}, \&
  {Turmon}}]{Hoeksema_etal_2014}
{Hoeksema}, J.~T., {Liu}, Y., {Hayashi}, K., {et~al.} 2014, \solphys, 289, 3483

\bibitem[{{Howard} {et~al.}(2008){Howard}, {Moses}, {Vourlidas}, {Newmark},
  {Socker}, {Plunkett}, {Korendyke}, {Cook}, {Hurley}, {Davila}, {Thompson},
  {St Cyr}, {Mentzell}, {Mehalick}, {Lemen}, {Wuelser}, {Duncan}, {Tarbell},
  {Wolfson}, {Moore}, {Harrison}, {Waltham}, {Lang}, {Davis}, {Eyles},
  {Mapson-Menard}, {Simnett}, {Halain}, {Defise}, {Mazy}, {Rochus}, {Mercier},
  {Ravet}, {Delmotte}, {Auchere}, {Delaboudiniere}, {Bothmer}, {Deutsch},
  {Wang}, {Rich}, {Cooper}, {Stephens}, {Maahs}, {Baugh}, {McMullin}, \&
  {Carter}}]{Howard_etal_2008}
{Howard}, R.~A., {Moses}, J.~D., {Vourlidas}, A., {et~al.} 2008, \ssr, 136, 67

\bibitem[{{Inoue} {et~al.}(2013){Inoue}, {Hayashi}, {Shiota}, {Magara}, \&
  {Choe}}]{Inoue_etal_2013}
{Inoue}, S., {Hayashi}, K., {Shiota}, D., {Magara}, T., \& {Choe}, G.~S. 2013,
  \apj, 770, 79

\bibitem[{{Inoue} {et~al.}(2011){Inoue}, {Kusano}, {Magara}, {Shiota}, \&
  {Yamamoto}}]{Inoue_etal_2011}
{Inoue}, S., {Kusano}, K., {Magara}, T., {Shiota}, D., \& {Yamamoto}, T.~T.
  2011, \apj, 738, 161

\bibitem[{{Inoue} {et~al.}(2012){Inoue}, {Shiota}, {Yamamoto}, {Pandey},
  {Magara}, \& {Choe}}]{Inoue_etal_2012}
{Inoue}, S., {Shiota}, D., {Yamamoto}, T.~T., {et~al.} 2012, \apj, 760, 17

\bibitem[{{Kaiser} {et~al.}(2008){Kaiser}, {Kucera}, {Davila}, {St.~Cyr},
  {Guhathakurta}, \& {Christian}}]{Kaiser_etal_2008}
{Kaiser}, M.~L., {Kucera}, T.~A., {Davila}, J.~M., {et~al.} 2008, \ssr, 136, 5

\bibitem[{{Kilpua} {et~al.}(2013){Kilpua}, {Isavnin}, {Vourlidas}, {Koskinen},
  \& {Rodriguez}}]{Kilpua_etal_2013}
{Kilpua}, E., {Isavnin}, A., {Vourlidas}, A., {Koskinen}, H., \& {Rodriguez},
  L. 2013, in EGU General Assembly Conference Abstracts, Vol.~15, EGU General
  Assembly Conference Abstracts, 2827

\bibitem[{{Kliem} \& {T{\"o}r{\"o}k}(2006)}]{Kliem_Torok_2006}
{Kliem}, B., \& {T{\"o}r{\"o}k}, T. 2006, Physical Review Letters, 96, 255002

\bibitem[{{Leake} {et~al.}(2013){Leake}, {Linton}, \&
  {T{\"o}r{\"o}k}}]{Leake_etal_2013}
{Leake}, J.~E., {Linton}, M.~G., \& {T{\"o}r{\"o}k}, T. 2013, \apj, 778, 99

\bibitem[{{Lemen} {et~al.}(2011){Lemen}, {Title}, {Akin}, {Boerner}, {Chou},
  {Drake}, {Duncan}, {Edwards}, {Friedlaender}, {Heyman}, {Hurlburt}, {Katz},
  {Kushner}, {Levay}, {Lindgren}, {Mathur}, {McFeaters}, {Mitchell}, {Rehse},
  {Schrijver}, {Springer}, {Stern}, {Tarbell}, {Wuelser}, {Wolfson}, {Yanari},
  {Bookbinder}, {Cheimets}, {Caldwell}, {Deluca}, {Gates}, {Golub}, {Park},
  {Podgorski}, {Bush}, {Scherrer}, {Gummin}, {Smith}, {Auker}, {Jerram},
  {Pool}, {Soufli}, {Windt}, {Beardsley}, {Clapp}, {Lang}, \&
  {Waltham}}]{Lemen_etal_2011}
{Lemen}, J.~R., {Title}, A.~M., {Akin}, D.~J., {et~al.} 2011, \solphys, 115

\bibitem[{{Lynch} {et~al.}(2008){Lynch}, {Antiochos}, {DeVore}, {Luhmann}, \&
  {Zurbuchen}}]{Lynch_etal_2008}
{Lynch}, B.~J., {Antiochos}, S.~K., {DeVore}, C.~R., {Luhmann}, J.~G., \&
  {Zurbuchen}, T.~H. 2008, \apj, 683, 1192

\bibitem[{{Lynch} {et~al.}(2005){Lynch}, {Gruesbeck}, {Zurbuchen}, \&
  {Antiochos}}]{Lynch_etal_2005}
{Lynch}, B.~J., {Gruesbeck}, J.~R., {Zurbuchen}, T.~H., \& {Antiochos}, S.~K.
  2005, Journal of Geophysical Research (Space Physics), 110, 8107

\bibitem[{{McKenzie} \& {Canfield}(2008)}]{McKenzie_Canfield_2008}
{McKenzie}, D.~E., \& {Canfield}, R.~C. 2008, \aap, 481, L65

\bibitem[{{M{\"o}stl} {et~al.}(2009){M{\"o}stl}, {Farrugia}, {Miklenic},
  {Temmer}, {Galvin}, {Luhmann}, {Kilpua}, {Leitner}, {Nieves-Chinchilla},
  {Veronig}, \& {Biernat}}]{Moestl_etal_2009}
{M{\"o}stl}, C., {Farrugia}, C.~J., {Miklenic}, C., {et~al.} 2009, Journal of
  Geophysical Research (Space Physics), 114, 4102

\bibitem[{Nindos {et~al.}(2015)Nindos, Patsourakos, Vourlidas, \&
  Tagikas}]{Nindos_etal_2015}
Nindos, A., Patsourakos, S., Vourlidas, A., \& Tagikas, C. 2015, Astrophys. J.,
  in print

\bibitem[{{Patsourakos} {et~al.}(2013){Patsourakos}, {Vourlidas}, \&
  {Stenborg}}]{Patsourakos_etal_2013}
{Patsourakos}, S., {Vourlidas}, A., \& {Stenborg}, G. 2013, \apj, 764, 125

\bibitem[{{Pesnell} {et~al.}(2012){Pesnell}, {Thompson}, \&
  {Chamberlin}}]{Pesnell_etal_2012}
{Pesnell}, W.~D., {Thompson}, B.~J., \& {Chamberlin}, P.~C. 2012, \solphys,
  275, 3

\bibitem[{{Petrie}(2012)}]{Petrie_2012}
{Petrie}, G.~J.~D. 2012, \apj, 759, 50

\bibitem[{{Plowman} {et~al.}(2013){Plowman}, {Kankelborg}, \&
  {Martens}}]{Plowman_etal_2013}
{Plowman}, J., {Kankelborg}, C., \& {Martens}, P. 2013, \apj, 771, 2

\bibitem[{{Qiu} {et~al.}(2007){Qiu}, {Hu}, {Howard}, \&
  {Yurchyshyn}}]{Qiu_etal_2007}
{Qiu}, J., {Hu}, Q., {Howard}, T.~A., \& {Yurchyshyn}, V.~B. 2007, \apj, 659,
  758

\bibitem[{{Richardson}(2013)}]{Richardson_2013}
{Richardson}, I.~G. 2013, Journal of Space Weather and Space Climate, 3, A8

\bibitem[{{Riley} \& {Richardson}(2013)}]{Riley_Richardson_2013}
{Riley}, P., \& {Richardson}, I.~G. 2013, \solphys, 284, 217

\bibitem[{{Rust} \& {Kumar}(1996)}]{Rust_Kumar_1996}
{Rust}, D.~M., \& {Kumar}, A. 1996, \apjl, 464, L199+

\bibitem[{{Savcheva} {et~al.}(2012){Savcheva}, {Pariat}, {van Ballegooijen},
  {Aulanier}, \& {DeLuca}}]{Savcheva_etal_2012}
{Savcheva}, A., {Pariat}, E., {van Ballegooijen}, A., {Aulanier}, G., \&
  {DeLuca}, E. 2012, \apj, 750, 15

\bibitem[{{Savcheva} \& {van Ballegooijen}(2009)}]{Savcheva_vanBallego_2009}
{Savcheva}, A., \& {van Ballegooijen}, A. 2009, \apj, 703, 1766

\bibitem[{{Scherrer} {et~al.}(2012){Scherrer}, {Schou}, {Bush}, {Kosovichev},
  {Bogart}, {Hoeksema}, {Liu}, {Duvall}, {Zhao}, {Title}, {Schrijver},
  {Tarbell}, \& {Tomczyk}}]{Scherrer_etal_2012}
{Scherrer}, P.~H., {Schou}, J., {Bush}, R.~I., {et~al.} 2012, \solphys, 275,
  207

\bibitem[{{Schrijver}(2007)}]{Schrijver_2007}
{Schrijver}, C.~J. 2007, \apjl, 655, L117

\bibitem[{{Su} {et~al.}(2009){Su}, {van Ballegooijen}, {Schmieder}, {Berlicki},
  {Guo}, {Golub}, \& {Huang}}]{Su_etal_2009}
{Su}, Y., {van Ballegooijen}, A., {Schmieder}, B., {et~al.} 2009, \apj, 704,
  341

\bibitem[{{Tanaka}(1991)}]{Tanaka_1991}
{Tanaka}, K. 1991, \solphys, 136, 133

\bibitem[{{T{\"o}r{\"o}k} \& {Kliem}(2005)}]{Torok_Kliem_2005}
{T{\"o}r{\"o}k}, T., \& {Kliem}, B. 2005, \apjl, 630, L97

\bibitem[{{Tziotziou} {et~al.}(2013){Tziotziou}, {Georgoulis}, \&
  {Liu}}]{Tziotziou_etal_2013}
{Tziotziou}, K., {Georgoulis}, M.~K., \& {Liu}, Y. 2013, \apj, 772, 115

\bibitem[{{van Ballegooijen}(2004)}]{vanBallegooijen_2004}
{van Ballegooijen}, A.~A. 2004, \apj, 612, 519

\bibitem[{{van Ballegooijen} \& {Martens}(1989)}]{vanBallegooijen_Martens_1989}
{van Ballegooijen}, A.~A., \& {Martens}, P.~C.~H. 1989, \apj, 343, 971

\bibitem[{{Vourlidas} {et~al.}(2013){Vourlidas}, {Lynch}, {Howard}, \&
  {Li}}]{Vourlidas_et_al_2013}
{Vourlidas}, A., {Lynch}, B.~J., {Howard}, R.~A., \& {Li}, Y. 2013, \solphys,
  284, 179

\bibitem[{{Wang} {et~al.}(1994){Wang}, {Ewell}, {Zirin}, \&
  {Ai}}]{Wang_etal_1994}
{Wang}, H., {Ewell}, Jr., M.~W., {Zirin}, H., \& {Ai}, G. 1994, \apj, 424, 436

\bibitem[{{Wheatland} {et~al.}(2000){Wheatland}, {Sturrock}, \&
  {Roumeliotis}}]{Wheatland_etal_2000}
{Wheatland}, M.~S., {Sturrock}, P.~A., \& {Roumeliotis}, G. 2000, \apj, 540,
  1150

\bibitem[{{Wiegelmann}(2004)}]{Wiegelmann_2004}
{Wiegelmann}, T. 2004, \solphys, 219, 87

\bibitem[{{Wiegelmann} {et~al.}(2006){Wiegelmann}, {Inhester}, \&
  {Sakurai}}]{Wiegelmann_etal_2006}
{Wiegelmann}, T., {Inhester}, B., \& {Sakurai}, T. 2006, \solphys, 233, 215

\bibitem[{{Wiegelmann} \& {Sakurai}(2012)}]{Wiegelmann_Sakurai_2012}
{Wiegelmann}, T., \& {Sakurai}, T. 2012, Living Reviews in Solar Physics, 9, 5

\bibitem[{{Wuelser} {et~al.}(2004){Wuelser}, {Lemen}, {Tarbell}, {Wolfson},
  {Cannon}, {Carpenter}, {Duncan}, {Gradwohl}, {Meyer}, {Moore}, {Navarro},
  {Pearson}, {Rossi}, {Springer}, {Howard}, {Moses}, {Newmark},
  {Delaboudiniere}, {Artzner}, {Auchere}, {Bougnet}, {Bouyries}, {Bridou},
  {Clotaire}, {Colas}, {Delmotte}, {Jerome}, {Lamare}, {Mercier}, {Mullot},
  {Ravet}, {Song}, {Bothmer}, \& {Deutsch}}]{Wuelser_etal_2004}
{Wuelser}, J.-P., {Lemen}, J.~R., {Tarbell}, T.~D., {et~al.} 2004, in Society
  of Photo-Optical Instrumentation Engineers (SPIE) Conference Series, Vol.
  5171, Telescopes and Instrumentation for Solar Astrophysics, ed.
  S.~{Fineschi} \& M.~A. {Gummin}, 111--122

\bibitem[{{Yurchyshyn} {et~al.}(2006){Yurchyshyn}, {Karlick{\'y}}, {Hu}, \&
  {Wang}}]{Yurchyshyn_etal_2006b}
{Yurchyshyn}, V., {Karlick{\'y}}, M., {Hu}, Q., \& {Wang}, H. 2006, \solphys,
  235, 147

\bibitem[{{Zhang} {et~al.}(2012){Zhang}, {Cheng}, \& {Ding}}]{Zhang_etal_2012}
{Zhang}, J., {Cheng}, X., \& {Ding}, M.-D. 2012, Nature Communications, 3, 747

\bibitem[{{Zhang} {et~al.}(2010){Zhang}, {Wang}, \& {Liu}}]{Zhang_etal_2010}
{Zhang}, J., {Wang}, Y., \& {Liu}, Y. 2010, \apj, 723, 1006

\bibitem[{{Zirin} \& {Tanaka}(1973)}]{Zirin_Tanaka_1973}
{Zirin}, H., \& {Tanaka}, K. 1973, \solphys, 32, 173

\end{thebibliography}

\clearpage

\clearpage

\clearpage

\clearpage

\clearpage

\clearpage

\clearpage

\clearpage

\clearpage

\clearpage

\clearpage

\clearpage

\clearpage

\clearpage

\clearpage

\clearpage

\begin{deluxetable}{ccccc}
\tablecolumns{8}
\tablewidth{0pc}
\tablecaption{MFR Length, Cross-section, Core Height and mean $|\mathbf{B}|$ before the CME eruptions\tablenotemark{*}}

\tablehead{
\colhead{MFR} & \colhead{Length (Mm)} & \colhead{Cross-section (Mm)\tablenotemark{\dagger}}& \colhead{Core height (Mm)\tablenotemark{\ddagger}} &\colhead{Mean  $|\mathbf{B}|$ (G)} } 
\startdata
NE MFR & 80 & 10 & 3  & 650\\ 
SW MFR & 40 & 6 & 3  & 500\\
\enddata
\tablenotetext{*}{Determined from the NLFFF extrapolation of March 6, 2012, 23:48~UT}
\tablenotetext{\dagger}{Cross-section inferred from the 20\% of peak $|\mathbf{J}|$ at the slice cut along the longest direction.}
\tablenotetext{\ddagger}{Height inferred from peak value of $|\mathbf{J}|$.}
\label{TABLE}
\end{deluxetable}


\appendix

\section{NLFFF Extrapolation}
The boundaries were binned twofold (2$\times$2 binning, i.e. down-sampled to 282$\times$195 pixels) to accommodate our computer resources. A preprocessing of these rebinned photospheric boundaries followed, to make them compatible with the force-free condition (as in \citealt{Wiegelmann_etal_2006}). The size of the computational domain is 282$\times$195$\times$118 pixels and a potential magnetic field extrapolation was carried for each boundary of our time-series (using the $B^{phot}_z$ component only) to serve as the initial field to be optimized into a NLFFF field. The potential extrapolations were calculated via a standard Green function integrator for the magnetic scalar potential, $\Phi_m$. Then the potential datacubes, derived as $\mathbf{B}_{potn}=-\nabla \Phi_m$, are fed into the NLFFF optimization routine along with the (preprocessed) photospheric boundaries ($B_x^{phot}$, $B_y^{phot}$,$B_z^{phot}$). The boundary layer at each of these faces was chosen to be 16 pixels thick. This resulted in a ``physical'' (i.e. numerical boundary-effect-free) sub-volume with a size of 250$\times$163$\times$106 pixels (or 180$\times$117$\times$76\,Mm$^{3}$) for each component of the NLFF magnetic field. This is the size of the final products of the NLFFF extrapolation used in our analysis.

\section{Twist Number Calculation}
Since points-of-avoidance or holes on the length-map signal overlapping field-line endpoints in  neighboring pixels, we modified the generation of length-maps in the following way. We increased  the sampling step for the seedpoints ten-fold using bi-linear interpolation and stored the locations of footpoints with the value for the length in a 10$\times$ larger length-map. This allowed for a 10-fold increase of the ``resolving power'' of the image grid for discriminating the locations of the end-points of field lines. This large length-map is therefore sparse and in order to fill the gaps we performed a forward integration from the locations of the holes at the positive polarity (end-points from the initial integration). Then the (now negative) end-points are stored in the same map and a Delaunay triangulation is performed. The latter is necessary in order to fill the remaining gaps within the negative and positive polarities via bi-linear interpolation. Then, multiplying this resulting length-map with its associated (i.e. 10$\times$ larger via bi-linear interpolation) $\bar{\alpha}$-map, over $4\pi$, yields the twist number, $T_n$, for the NLFFF extrapolated field lines in the physical domain. In this way we can derive smooth and continuous $T_n$ maps for each of the NLFFF datacubes and now can study the time evolution of field-line twist as a way to monitor topological changes in the NLFFF.

\end{document}